\documentclass{aa}  
\usepackage{graphicx}
\usepackage{txfonts}
\usepackage{subcaption}         
\usepackage{lscape}             
\usepackage{placeins}           
\usepackage{natbib}
\bibpunct{(}{)}{;}{a}{}{,}             

\usepackage{amsmath}
\usepackage{enumitem} 
\usepackage[colorlinks=True]{hyperref}
\hypersetup{
    linkcolor=blue,  
    citecolor=blue,
    }
\usepackage{graphicx}
\usepackage{booktabs}
\usepackage{txfonts}
\DeclareUnicodeCharacter{2212}{-}
\DeclareUnicodeCharacter{2061}{}
\begin{document}

   \title{The S-PLUS Fornax Project (S+FP): An extragalactic catalog covering $\sim$ 5 virial radii around NGC\,1399 with galaxy properties}


   \author{R. F. Haack
          \inst{1}\fnmsep\inst{2}\fnmsep\inst{3},
          A. V. Smith Castelli\inst{1}\fnmsep\inst{2}\fnmsep\inst{3},
          L. Sodré Jr.\inst{3},
          C. Mendes de Oliveira\inst{3},
          A. R. Lopes\inst{1},
          L. A. Gutiérrez-Soto\inst{1},
          R. Demarco\inst{4},
          D. E. Olave-Rojas\inst{5},
          E. R. Carrasco\inst{6},
          P. K. Humire\inst{3},
          J. P. Calderón\inst{1}\fnmsep\inst{2},
          F. de Almeida Fernandes\inst{7},
          \\
          L. Lomel\'i-N\'u\~nez\inst{8},
          G. Sepúlveda\inst{5},
          C. Lima-Dias\inst{9},
          S. Torres Flores\inst{9},
          E. Telles\inst{10},
          N. M. Cardoso\inst{3},
          D. Palma\inst{3},
          \\
          L. Doubrawa\inst{3},
          D. Pallero\inst{11}\fnmsep\inst{12},
          M. Marinello\inst{13},
          W. Schoenell\inst{14},
          T. Ribeiro\inst{15}
          \and
          A. Kanaan\inst{16}
          }

   \institute{Instituto de Astrofísica de La Plata, UNLP-CONICET,  Paseo del Bosque s/n, La Plata, B1900FWA, Argentina
         \and
             Facultad\,de\,Ciencias\,Astronómicas\,y\,Geofísicas,\,Universidad\,Nacional\,de\,La Plata,\,Paseo\,del\,Bosque\,s/n,\,La\,Plata,\,B1900FWA,\,Argentina
        \and
             Departamento de Astronomia, Instituto de Astronomia, Geofísica e Ciências Atmosféricas da USP, Cidade Universitária, 05508-090 São Paulo, SP, Brazil
        \and
             Institute of Astrophysics, Facultad de Ciencias Exactas, Universidad Andrés Bello, Sede Concepción, Talcahuano, Chile
        \and
             Departamento de Tecnologías Industriales, Facultad de Ingeniería, Universidad de Talca, Los Niches km 1, Curicó, Chile
        \and
            International Gemini Observatory/NSF's National Optical-Infrared Research Laboratory, Casilla 603, La Serena, Chile
        \and
            Universidade do Vale do Paraíba, Av. Shishima Hifumi, 2911, São José dos Campos, SP, 12244-000, Brazil
        \and
            Valongo\,Observatory,\,Federal\,University\,of\,Rio\,de Janeiro,\,Ladeira\,Pedro\,Antonio\,43,\,Saude\,Rio\,de\,Janeiro,\,RJ, 20080-090,\,Brazil
        \and 
            Departamento de Astronomía, Universidad de La Serena, Av. J. Cisternas 1200 N, 1720236 La Serena, Chile
        \and
            Observatório Nacional, Rua General José Cristino, 77, São Cristóvão, 20921-400 Rio de Janeiro, RJ, Brazil
        \and
            Departamento de Física, Universidad Técnica Federico Santa María, Avenida España 1680, Valparaíso, Chile
        \and
            Millennium Nucleus for Galaxies (MINGAL)
        \and
            Laboratório Nacional de Astrofísica, Rua Estados Unidos 154, Itajubá, 37504-364, MG, Brazil
        \and
            GMTO Corporation 465 N. Halstead Street, Suite 250 Pasadena, CA 91107, USA
        \and
            Rubin Observatory Project Office, 950 N. Cherry Ave., Tucson, AZ 85719, USA
        \and
            Departamento de Física - CFM - Universidade Federal de Santa Catarina, PO BOx 476, 88040-900, Florianópolis, SC, Brazil
             }

   \date{Received September 15, 1996; accepted March 16, 1997}

 
  \abstract
   {
   Observational extragalactic catalogs over wide sky areas are essential for uncovering the large-scale structure of the Universe. They allow, among others, cosmological studies and density analyses that impose strong constraints on models of galaxy formation and evolution.
   }
   {By taking advantage of the wide field images and the 12 optical bands of the Southern Photometric Local Universe Survey (S-PLUS), we aim at providing a catalog of galaxies located, in projection, towards the Fornax galaxy cluster, within $\sim$ 5 virial radii in right ascension (R.A.) and $\sim$ 3 virial radius in declination (Dec) around NGC\,1399, the dominant galaxy of the cluster. Such a catalog will allow unprecedented large-scale structure studies in that sky region.
   }
   {Supervised deep learning algorithms have been developed, utilizing neural networks complemented with dimensionality reduction techniques, to classify and separate spurious objects, stars and galaxies in a photometric catalog previously built for the S-PLUS Fornax Project (S+FP). That catalog was built using a combination of SExtractor configurations optimized for galaxy detection and characterization.
   }
   {A catalog of 119,580 galaxies was obtained in the direction of the Fornax cluster  containing photometric information in the 12 optical bands of S-PLUS complemented with GALEX (UV), VHS-VISTA (NIR) and AllWISE (MIR) data. We estimate photometric redshifts ($\sigma_{\mathrm{NMAD}} \sim 0.0219$) with a lower limit of $z_{\mathrm{lim}}$ $\sim$ 0.03. Stellar masses, star formation rates (SFRs) and $D4000_{N}$ index estimates were obtained through a machine learning approach, by matching S-PLUS photometric data to SDSS spectroscopic data. The completeness of the catalog (72\%) was calculated by comparing with mock catalogs.
   } 
   {Taking into account our $z_{\mathrm{lim}}$, we were able to identify 119,230  background galaxies and to find 350 candidates to be Fornax members or infalling galaxies, not included in our compilation of 1,005 galaxies previously reported in the literature. We were also able to classify the galaxies in our catalog as quiescent ($43\%$), star-forming ($39\%$) and transition  ($18\%$) galaxies. In addition, 181 emission line galaxy (ELG) candidates were identified using the filter J0660.  
   The spatial distribution of the galaxies in our catalog show projected overdensities that match 89 background clusters identified by eROSITA. This confirms the robustness of our catalog in tracing real structures. In that context, we expect that the extragalactic catalog of the S+FP allows us to better understand the large-scale structure in the direction of the Fornax cluster, and to identify the substructures that are feeding Fornax.}

\titlerunning{The Extragalactic Catalog of the S+FP}
\authorrunning{Haack et al.}
   \maketitle
%

\section{Introduction}
Large photometric extragalactic catalogs constitute a basic and straightforward tool for revealing the projected spatial distribution of the galaxy distribution in the sky and, as a consequence, for disclosing the large-scale structure of the Universe. Combined with additional information such as that provided by a density analysis and spectroscopic data, among others, they allow a number of studies in the cosmological field to explore the processes of galaxy group and cluster formation, and to deepen the comprehension of the formation and evolution paths of the galaxies belonging to different environments.


In recent years, several photometric surveys have allowed the creation of large extragalactic catalogs. The Sloan Digital Sky Survey (SDSS; \cite{York2000}) revolutionized observational astronomy with its imaging coverage of $\sim$14,000 deg² of the northern sky in five optical broad bands (u,g,r,i,z). The Dark Energy Camera Legacy Survey (DECaLS; \citealt{Dey2019})is performing a deeper mapping of the southern hemisphere, through wide-field images in four optical bands (g,r,i,z). The Wide-field Infrared Survey Explorer (WISE; \citealt{Wright2010}) and the Vista Hemisphere Survey (VHS; \citealt{McMahon2013}) have complemented that optical data in the IR regime. For objects simultaneously included in those catalogs, the combination of photometry at different wavelenghts opens the possibility of obtaining additional valuable information, like photometric redshifts ($z_{\mathrm{phot}}$)  and stellar masses, from the fit of their spectral energy distribution (SED). However, the combination of just broad bands (such as those in the SDSS) brings limitations in estimating those parameters. To improve this issue, projects such as the Southern Photometric Local Universe Survey (S-PLUS; \citealp{MendesdeOliveira2019}) and its northern counterpart, the Javalambre Photometric Local Universe Survey (J-PLUS; \citealp{Cenarro2019}), have implemented the Javalambre photometric system consisting of 12 optical broad and narrow bands. J-PAS \citep{Benitez2015} has gone further by using 56 narrow bands in the optical range, which allows for quasi-spectroscopic sampling of the continuum and emission lines of galaxies.

A fundamental aspect in the construction of a large extragalactic catalog, from automatic photometry performed in wide-field images using software like SExtractor \citep{Bertin1996}, is the separation of galaxies from compact sources and spurious detections in a confident manner. Several works have addressed this problem using morphological and photometric criteria, as well as through the application of machine learning techniques. As an example of the latter approach, \cite{Bailer-Jones2019} have used neural networks and probabilistic models to perform a galaxy-quasar separation in Gaia Data Release 2.

Once a reliable multiband catalog of galaxies is obtained, the estimation of $z_{\mathrm{phot}}$ becomes a key property in order to trace large-scale structures. In general, those $z_{\mathrm{phot}}$ can be obtained in two ways: via SED-fitting or applying machine learning (ML) techniques. SED-fitting methods infer the redshift of a galaxy by comparing its observed photometry with stellar population models. These models can be observational, synthetic, or a combination of both. There are numerous codes that implement this approach, including LePhare \citep{Arnouts1999,Ilbert2006}, EAZY \citep{Brammer2008}, BAGPIPES \citep{Carnall2018}, CIGALE \citep{CIGALE} and AlStar \citep{Thaina2023}, which is an adaptation of the spectral fitting code STARLIGHT \citep{CidFernandes2005}. Several studies have compared the accuracy for some of these codes under different observational conditions (see, e.g., \citealt{Dahlen2013,Schmidt2020,Humire2025}).

Photometric redshifts can also be estimated using ML algorithms, which have experienced a rapid development over the last decade due to their ability to model complex nonlinear relationships in high-dimensional data. These methods include random forests, support vector machines, Gaussian processes, and deep neural networks \citep[]{Cavuoti2017,Pasquet2019,Zhou2021,Lima2022,Teixeira2024}. Unlike SED-fitting, which relies on physical models, ML approaches are purely empirical and require a representative training set with known spectroscopic redshifts. Once trained, these models can predict $z_{\mathrm{phot}}$ with high computational efficiency, making them attractive for large datasets.

An important distinction between the two approaches lies in their performance across different redshift regimes. ML methods tend to display a good performance at low redshifts ($z \lesssim 1$), particularly when the training set is sufficiently dense and covers the relevant color space. However, their performance degrades at higher redshifts or in regions of parameter space not well represented in the training sample. In contrast, SED-fitting methods, though typically slower and more sensitive to photometric uncertainties and template mismatches, are more robust in exploring regions of parameter space where models can be physically extrapolated and can reach higher redshifts \citep[]{Beck2017,Duncan2018}.

Each approach has its own set of advantages and limitations. SED-fitting provides not only redshift estimates but also physical properties such as stellar masses, SFRs, and extinction values, derived from the same model. However, it highly depends on the choice of templates, priors, and assumptions about stellar population synthesis, which can introduce systematics. ML methods, by contrast, are more flexible and often achieve a lower scatter and lower outlier rates in well-calibrated regimes, but they may lack interpretability and are generally less suited for extrapolation. As a result, hybrid approaches that combine the strengths of both methods have also been proposed \citep[]{DIsanto2018,Schmidt2020}.

Besides redshift estimation, one of the most crucial parameters that an extragalactic catalog may provide is the galaxy stellar mass. Stellar mass is a fundamental quantity for understanding galaxy evolution, as it correlates with various physical properties such as star formation rate (SFR), metallicity, and morphology \citep{Gallazzi2005,Peng2010}. However, even when spectroscopic redshifts are available, the estimation of stellar masses is not straightforward. It typically involves fitting the galaxy's SED with stellar population synthesis models, which require assumptions about the initial mass function (IMF), dust attenuation, and star formation history. Variations in these assumptions can lead to systematic uncertainties of up to 0.3 dex in mass values \citep{Conroy2013,Pacifici2023}. In photometric surveys, where redshift uncertainties propagate into the mass estimates, the challenges are even greater. Accurate redshifts are essential for robust mass determinations, especially for faint or high-redshift galaxies.

In addition, the identification of emission line galaxies (ELGs), e.g. [OII], H$\alpha$ and Lyman-$\alpha$ emitters, is of particular interest. These sources are often associated with active star formation or nuclear activity, and can serve as tracers of the cosmic web and large-scale structures \citep{Cochrane2018,Khostovan2020}. Their strong emission lines make them easy to detect and characterize, even in low signal-to-noise data, and their spatial distribution over large scales can reveal the underlying matter density field. Therefore, the combination of accurate $z_{\mathrm{phot}}$, stellar mass estimates, and emission-line diagnostics is key to build comprehensive extragalactic catalogs that enable statistical studies of galaxy evolution and cosmology.

The Fornax cluster, located at a distance of $\sim$20 Mpc (z$\sim$0.0046; \citealt{Blakeslee2009}), is the closest rich galaxy cluster after the Virgo cluster. Fornax is particularly interesting for studies of galaxy formation and evolution because of its dynamic structure, with a central concentration dominated by NGC\,1399 and a significant population of dwarf galaxies. It also has a secondary substructure centered on NGC\,1316 (Fornax\,A) which is falling towards the main structure \citep{Scharf2005,Venhola2019}. Furthermore, the Fornax cluster  is part of the Eridanus-Fornax-Doradus complex. This large-scale structure was first identified in the context of the Southern Sky Redshift Survey  by \cite{daCosta1998}. According to these authors, the two most prominent structures in the $0 < v < 3000$ kms$^{-1}$ window, the Fornax cluster  and the Eridanus group \citep{Raj2024}, appear to form a linear structure connected to the looser Dorado group \citep{Kilborn2005}. 

The present work is developed in the context of the S-PLUS Fornax Project (S+FP; \citealp{SmithCastelli2024}), an initiative aimed at studying the Fornax galaxy cluster and its surroundings in 12 optical bands up to 5 virial radii ($R_{vir}$) in Right Ascension (R.A.) and $\sim$ 3 $R_{vir}$ in Declination (Dec.). In this paper we aim at presenting an extragalactic catalog that allows us to analyze the large scale distribution of galaxies in a projected sky area of $\sim 208$ deg$^2$ around NGC\,1399. To that aim, we estimate $z_{\mathrm{phot}}$, stellar masses, SFR, $D4000_{N}$ index and identify galaxies displaying an excess in the J0660 filter of S-PLUS that can be considered as emission line galaxy (ELG) candidates. 

This paper is structured as follows. Section\,\ref{sec:data} describes the photometric and spectroscopic data used. Section\,\ref{sec:methodology} details the methods of object classification and cleaning, including the separation between stars, galaxies and spurious objects. Section\,\ref{sec:properties} presents the estimation of $z_{\mathrm{phot}}$, stellar masses, SFRs and $D4000_{N}$ index values, an assessment of the accuracy of the methods used, and a lower limit in the estimation of $z_{\mathrm{phot}}$. Finally, Section\,\ref{sec:conclusions} summarizes our results, presents the conclusions and discusses possible future applications of this catalog.

\section{Data}
\label{sec:data}
\subsection{S-PLUS}
The Southern Photometric Local Universe Survey (S-PLUS; \citealt{MendesdeOliveira2019}) aims to map over 9,000 deg$^{2}$ of the southern sky using an 80-cm robotic telescope located at Cerro Tololo, Chile, equipped with a 12-band filter system. The uniqueness of S-PLUS is due to the use of seven narrowband filters (J0378, J0395 J0410, J0430, J0515, J0660, and J0861; \citealt{Cenarro2019}), developed specifically to probe interesting emission and absorption lines or bands in the nearby Universe such as [OII] 3727,3729, Ca H + K, H$\delta$, G-band, Mgb triplet, H$\alpha$, and Ca triplet.

In this work, we have used the photometry presented by \citet{Haack2024}, optimized to properly detect and characterize extragalactic objects. Our base catalog consists of a combination of the three catalogs, RUN\,1, RUN\,2 and RUN\,3, obtained in that work, avoiding the duplication of objects. It includes $\sim3\times10^{6}$ detected sources in a sky-projected area of $\sim$208 deg$^2$ around NGC\,1399, the dominant galaxy of the Fornax cluster. Those sources comprise galaxies, stars, and spurious objects. RUN\,1 detects and performs the photometry for faint and small galaxies, especially those near bright galaxies, RUN\,2 characterizes galaxies that are intermediate in both brightness and size, and RUN\,3 detects the largest galaxies, without subdividing them into several sources. 

In the context of the S+FP, a reference sample of 1,005 Fornax galaxies reported in the literature as spectroscopically confirmed members or probable members according to morphological criteria, has been established (for details on the compilation, we refer the reader to \citealt{SmithCastelli2024}). Hereafter, we will refer to this sample as the Fornax Literature Sample (FLS).

\subsection{Complementary photometric and spectroscopic data} 

To define the training sample for performing a star/galaxy separation, we use the probability for a source to be a star, a galaxy or a quasar from GAIA DR3 \citep{GAIA2023}.

In order to estimate $z_{\mathrm{phot}}$, we complement the photometry provided by S-PLUS with magnitudes in the UV from GALEX \citep{Bianchi2017}, NIR from VHS DR5 \citep{McMahon2013} and MIR from AllWISE \citep{Cutri2013}. The GALEX survey probed the entire sky using two UV bands, NUV and FUV, with effective wavelengths of 2315.7\textup{~\AA} and 1538.6\textup{~\AA}, respectively, and typical depths of 19.9 and 20.8 AB mag. VHS observed the southern hemisphere using four NIR bands, namely Y (0.88 $\mu$m), J (1.03 $\mu$m), H (1.25 $\mu$m) and Ks (2.20 $\mu$m), with coverages of 4825 deg$^{2}$, 16,689 deg$^{2}$, 2901 deg$^{2}$, and 16,684 deg$^{2}$, respectively, and 5$\sigma$ depths of 21.1, 20.8, 20.5, and 20.0 mag. This project observed more than one billion sources. WISE \citep{Wright2010} is a NASA Medium Class Explorer mission that conducted a digital imaging survey of the entire sky in the 3.4, 4.6, 12 and 22 $\mu$m MIR bandpasses (hereafter W1, W2, W3 and W4). The AllWISE program extends the work of the successful WISE mission by combining data from the cryogenic and post-cryogenic survey phases to provide the most comprehensive view of the MIR sky currently available. Also in Section\,\ref{sec:zphot}, 
we used spectroscopic redshifts collected by \citet{Lima2022} and obtained from SIMBAD to validate our estimations of $z_{\mathrm{phot}}$. In Section\,\ref{sec:masses} and Section\,\ref{sec:SFR-D4000} we made use of data from SDSS DR8 \citep{Aihara2011} and physical parameters from \cite{Kauffmann2003}, \cite{Brinchmann2004} and \cite{Tremonti2004} to estimate, through a ML approach, $z_{\mathrm{phot}}$, stellar masses, SFRs and $D4000_{N}$ index values for all the objects in the S+FP extragalactic catalog.

\section{Methodology}
\label{sec:methodology}

\subsection{Spurious objects identification}
\label{sec:spurious_identification}
As a starting point, each of the RUN\,1, RUN\,2 and RUN\,3 catalogs \citep{Haack2024} was taken separately and cleaned by internal duplications, performing a cross-match in R.A. and Dec. coordinates with a 3-arcsec error. This corresponds to $\sim$ 6 pixels of separation in the S-PLUS images. 

\begin{figure}[h!]
    \centering
    \includegraphics[width=1\columnwidth]{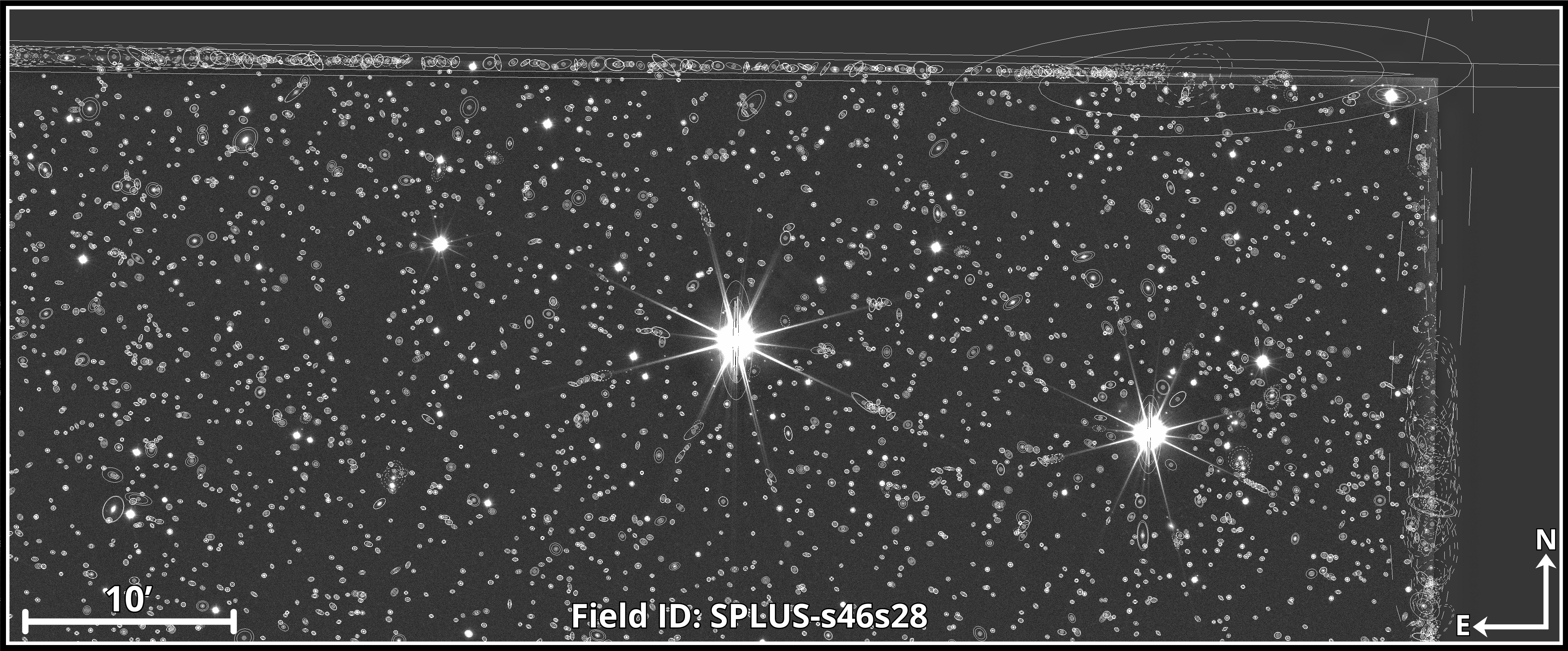}
    \caption{Definition of two categories of spurious objects: those detected close to the spikes of saturated star and those accumulated at the edges of the images.}
    \label{fig:def_espureos}
\end{figure}

In order to separate spurious objects from galaxies and stars (non-spurious objects), we have defined spurious detections in two categories: those arising at the edges of the images and those detected on the extended spikes of saturated stars. The S-PLUS DR4 images have lines or columns with null signal values over several pixels at their edges. In those regions, the mesh used by SExtractor for sky estimation finds null and non-null values which results in a wrong estimation of the sky level. On the other hand, spurious objects that accumulate very close to the spikes of saturated stars, especially the brightest ones, arise because the signal from the pattern of the spikes prevents  an efficient background assessment by SExtractor. Both cases are illustrated in Figure\,\ref{fig:def_espureos}. 

Once the spurious objects were defined, a supervised classification algorithm was developed using neural networks (NN) to separate non-spurious and spurious detections. The NN architecture used consists of three hidden layers (256, 128 and 64 neurons) with ReLU activation \citep{Nair2010}, and L2 regularization (0.001) to mitigate overfitting \citep{Ng2004}. Dropout (0.3) is applied to improve generalization and prevent excessive co-adaptation of neurons \citep{Srivastava2014}. The output layer uses a softmax function with two units for binary classification, employing a sparse categorical cross-entropy as a loss function, and Adam as an optimizer \citep{Kingma2017}. 

The learning of the NN to perform the classification is based on the photospectra of each of the sources, which is a discrete sequence of multiband magnitudes or fluxes that approximate the SED of an astrophysical source. Specifically, the input of the algorithm consists of the 12 AUTO (Kron-like; \citealp{Kron1980}) magnitudes and their respective errors. Examples of different photospectra corresponding to a quasar, a main-sequence star, an early-type galaxy, a planetary nebula, and a symbiotic star, can be seen in figure 7 of \cite{MendesdeOliveira2019}. 

To define the training and testing samples, both types of spurious objects have been manually labeled, rigorously checking each of the objects. 
In addition, we labeled stars 
(CLASS\_STAR\_r > 0.95) and galaxies (CLASS\_STAR\_r < 0.35) with r-band magnitudes below 21.3, taking into account that the labeled objects are of different sizes, brightnesses and colors, as well as of different morphologies in the case of galaxies. CLASS\_STAR is the stellarity index given by SExtractor. Stars and galaxies constitute the non-spurious category. In this way, our training and testing samples include 3,427 spurious objects (half of them are detections at the edges of the images and the other half corresponds to identifications near stars spikes) and 6,854 non-spurious objects (including stars and galaxies). Since the sample is not balanced between these two categories, the NN classification adopts Stratified K-Fold Cross-validation \citep{kohavi1995study}, dividing the data into three folds while maintaining the original proportion of classes, assigning greater weight to minority classes during training and preventing any fold from having zero representation of minority classes. The samples are separated into training (80$\%$) and testing (20$\%$) sets.

\begin{figure}[h!]
    \centering
    \includegraphics[width=1\columnwidth]{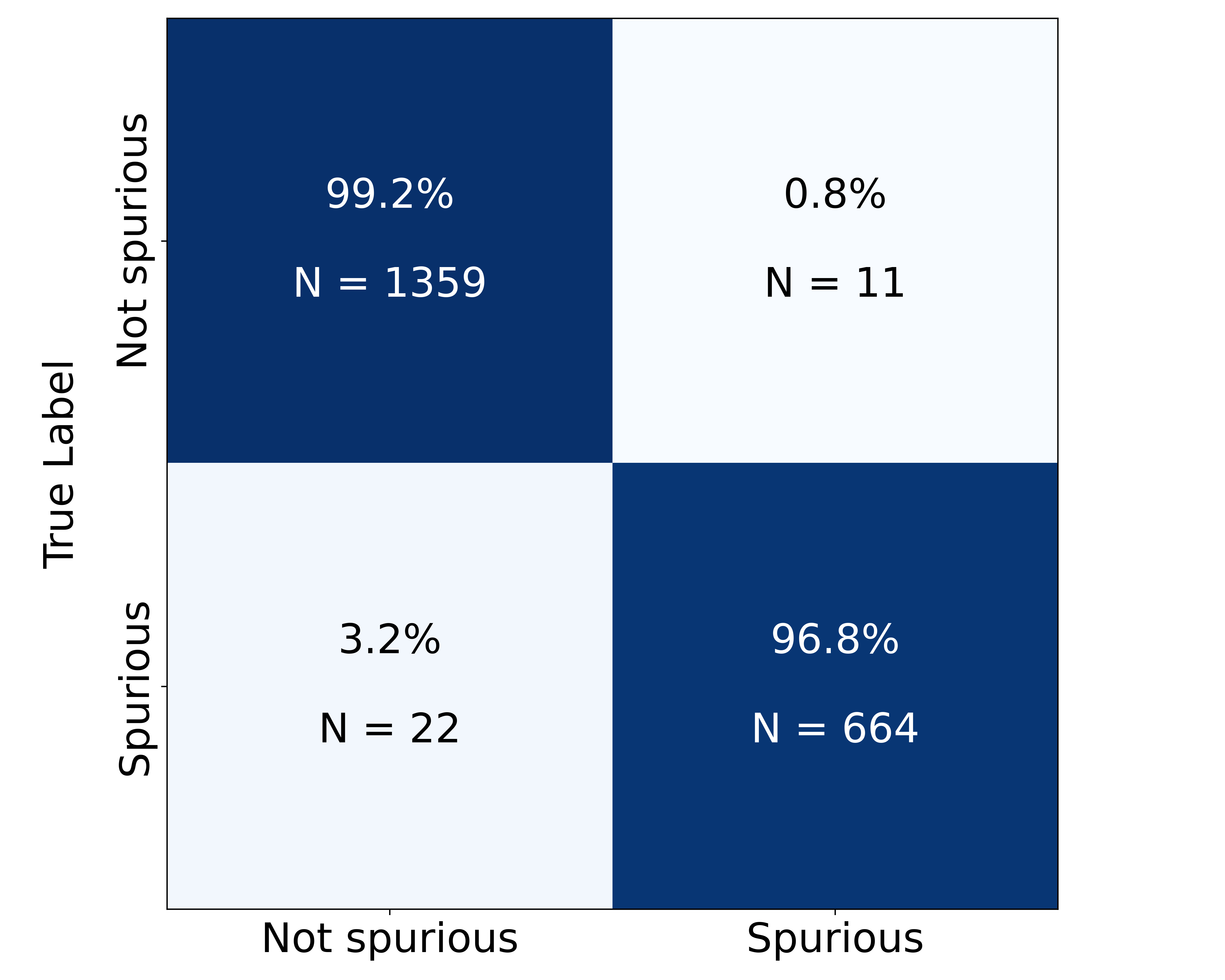}
    \caption{Confusion matrix for the separation of spurious and non-spurious objects for the test sample (20$\%$). On the vertical axis we show the true source labels and on the horizontal axis, the predicted labels. N represents the number of sources per label.}
    \label{fig:1_matriz}
\end{figure}

The confusion matrix of the best model can be seen in Figure\,\ref{fig:1_matriz}.
A very good classification has been achieved for the separation between spurious and non-spurious objects.
Only 3.2\% of the spurious detections are wrongly classified as non-spurious, while 0.8\% of non-spurious objects are uncorrectly classified as spurious detections.  
Therefore, we applied this model to the RUN\,1 and RUN\,2 catalogs in order to clean them from spurious objects. We should stress that, by construction, in the RUN\,3 catalog there are no spurious objects of the types previously defined, as the configuration file of this RUN was optimized to detect only large and bright objects.

As a next step, a merge of the three catalogs is made following a hierarchical assembly, since there are objects that appear in two or even three catalogs with differences in the size of the detections. Objects within a projected distance less than 3 arcsec were considered duplicates. To build the spurious-cleaned catalog, the fusion process prioritizes the object's entry from the RUN with the largest detection size. In that way, the process guarantees that the largest galaxies are included in the spurious-cleaned catalogue as a single object and not subdivided into several smaller sources. Briefly, the objects that appear in RUN\,1 and RUN\,2 will be characterized by the astrometric and photometric information obtained by RUN\,2. Those that were detected by RUN\,2 and RUN\,3 will be characterized by the information provided by RUN\,3. And those that were detected by RUN\,1, RUN\,2 and RUN\,3 will be included with the coordinates and photometry obtained from RUN\,3. In that sense, RUN\,1 will only characterize the objects that were only detected by RUN\,1. After this, we have a combined spurious-cleaned catalogue of 404,487 non-duplicated objects. 

\subsection{Star/Galaxy separation}
\label{sec:star-galaxy}

The CLASS\_STAR parameter provided by SExtractor, commonly used to separate resolved and unresolved sources, presents a classification ambiguity for faint and compact objects (figure 9 of \citealp[]{Bertin1996}). In order to avoid as much as possible such ambiguity, we decided to perform a better star/galaxy separation using Deep Learning (DL) algorithms with the same architecture explained in Section\,\ref{sec:spurious_identification}. 

For labeling the objects in the training and test samples, we took into account the information from GAIA \citep{GAIA2023}. By cross-matching the 404,487 sources included in the spurious-cleaned catalog with GAIA DR3 and considering a 1-arcsec error, we obtain a new catalog with 295,639 objects. According to the GAIA classification of sources, the latter is now separated into four categories: Star, Galaxy, QSO and Unknown. Based on the probability of confidence (P) of each class also provided by GAIA, we will consider this classification as reliable for stars, galaxies and QSOs if P$_{Star}$ $\geq$ 0.95, P$_{Galaxy}$ $\geq$ 0.95 and P$_{QSO}$ $\geq$ 0.95, respectively, where P$_{Star}$ is the probability given by GAIA to be a star, P$_{Galaxy}$ is the probability to be a galaxy, and P$_{QSO}$ is the probability to be a QSO. We will consider as Unknown (i.e. unreliable) sources all objects that present, simultaneously, P$_{Star}$ < 0.95 and P$_{Galaxy}$ < 0.95 and P$_{QSO}$ < 0.95. A similar strategy was adopted by \cite{Nakazono2021}. In our case, the selection made by GAIA was complemented with SDSS DR16 data only for sources with r > 18 mag, with the aim of increasing the number of both stars and galaxies at the faint end. This is necessary because GAIA's photometric depth is limited to r $\sim$ 18 mag. To do this, an S-PLUS catalog of the Stripe-82 region was cross-matched with SDSS data. This catalog contains photometry similar to that used in the Fornax direction (see \citealp{Haack2024}). Therefore, using it as a pivot catalog ensures consistency. Those galaxies attached for r > 18 mag have $z_{spec}$ > 0.002 and stars have $z_{spec}$ < 0.002, as measured in SDSS DR16.

For the training sample, we consider 13,187 stars and 12,786 galaxies with a GAIA+SDSS confident classification and also balanced between RUNs. Once again, 20$\%$ of each category was separated for the test sample and it was not used for training. Since the classes are balanced, stratification is not necessary and the model evaluation can be performed by standard cross-validation. This approach allows for a more efficient use of the data, especially in contexts where the distribution of classes is equal. We analyzed the learning history for both loss and accuracy metrics, implementing early stopping with a patience of 10 epochs, that is, allowing the model to train for up to 10 additional epochs without improvement before stopping. The convergence of training and validation curves demonstrates stable learning without divergence, indicating no evidence of overfitting in the final model. This controlled training approach ensures optimal generalization while preventing model deterioration.

For the application sample, that is, objects that we want to classify with the trained DL model, we take all Unknown objects from GAIA besides those sources that were lost in the cross-match with GAIA (78,523 sources). Our first model learns over the 66 colors corresponding to the 12 AUTO magnitudes of S-PLUS and geometric parameters that account for the size and concentration of the sources. The second model reduces the dimensionality of the data in a linear way by applying a Principal Component Analysis (PCA; \citealt{Bishop2006}). From this approach, we obtain that the 19 first components are enough to reach 99$\%$ of the total variance.
Our third model reduces the dimensionality of the data in a nonlinear way using Uniform Manifold Approximation and Projection for Dimension Reduction (UMAP; \citealt{McInnes2018}), 
arriving at 6 components to reach 99$\%$ of the total variance. Compared to the 66-color model and the UMAP model, the application of PCA turns out to be the model with the best accuracy and F1-score \citep{Rijsbergen1979}. The F1-score is defined as the harmonic mean of precision and recall, providing a balanced measure of performance. This statistics are shown in Table\,\ref{tab:f1-score}. 

\begin{table}[ht]
\centering
\caption{Comparison of accuracy and F1-Score for the 66 colors, PCA and UMAP models.}
\begin{tabular}{c|c|c|c}
 & 66 colors & PCA & UMAP \\
\hline
Accuracy & 0.951 & 0.970 & 0.943 \\
\hline
F1-Score & 0.950 & 0.969 & 0.942 \\
\end{tabular}
\label{tab:f1-score}
\end{table}

\begin{figure}[h!]
    \centering
    \includegraphics[width=0.85\columnwidth]{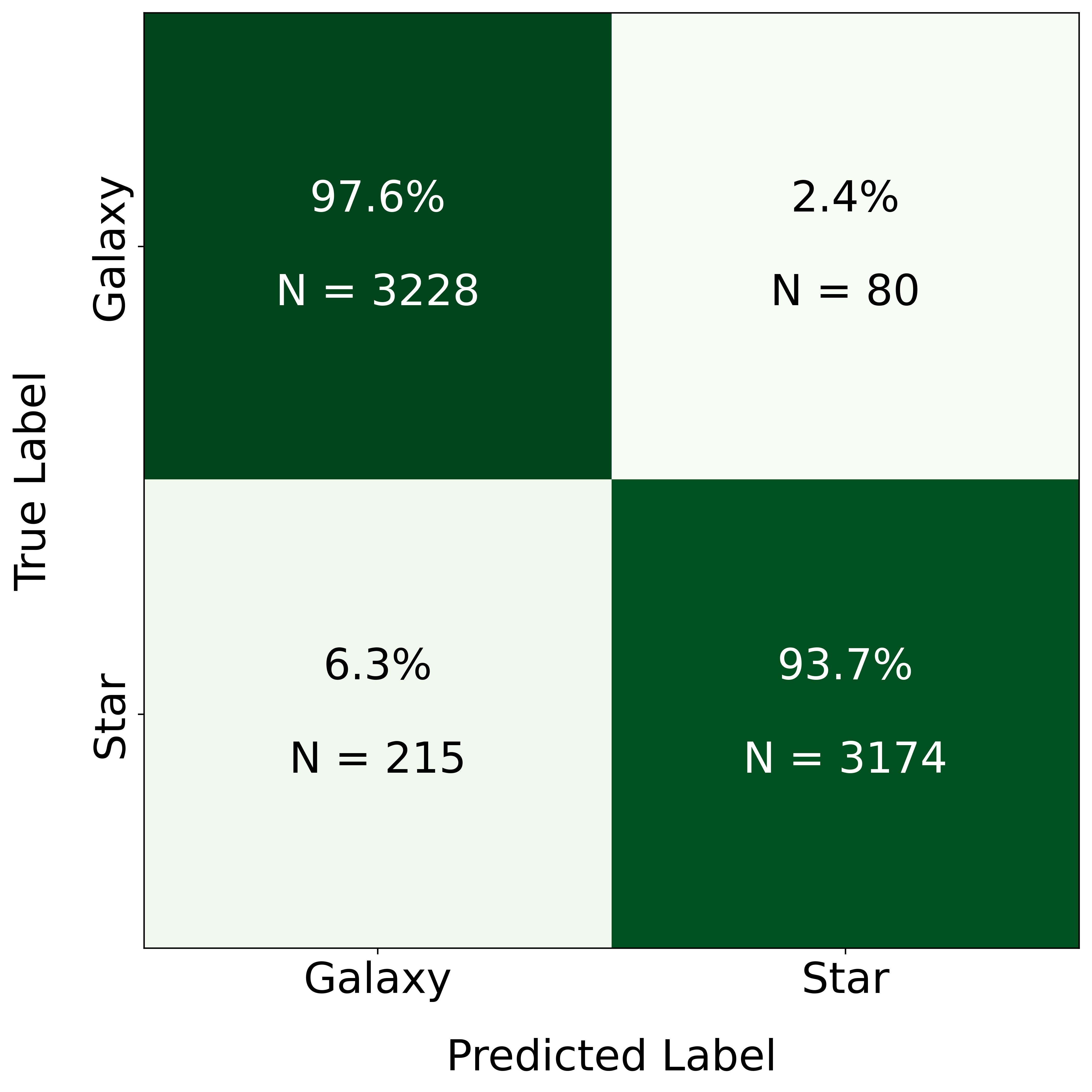}
    \caption{Confusion matrix of the PCA model, the one with the highest accuracy and F1-score. N represents the number of sources per label and the quantities shown correspond to the test sample (20$\%$) of the training sample.}
    \label{fig:matrices}
\end{figure}

The F1-score and accuracy are key metrics in the evaluation of classification models. Accuracy indicates how many of the model's positive predictions are actually correct, which is useful for minimizing false positives. On the other hand, the F1-score combines accuracy and the model's ability to correctly identify positive instances (recall), providing a balanced metric of the classifier performance. In a balanced data set, such as the one used in this study, the F1-score remains relevant to measure the overall performance without bias towards any of the classes. Since none of the classes dominates over the others, this metric acts as a clear benchmark of the effectiveness of the model, ensuring that the performance is not solely dependent on the accuracy or sensitivity of the classifier \citep{SOKOLOVA2009,Powers2020}. Figure\,\ref{fig:matrices} shows the confusion matrix of the PCA model.

Applying the PCA model to our sample, among the 404,487 sources,  269,894  were classified as stars and 134,593 as galaxies. The analysis of the characteristics that contribute the most to each component of the PCA, correlations and anticorrelations, is explained in Appendix\,\ref{app:histoPCA}. 

\begin{figure*}[h!]
    \centering
    \includegraphics[width=1\textwidth]{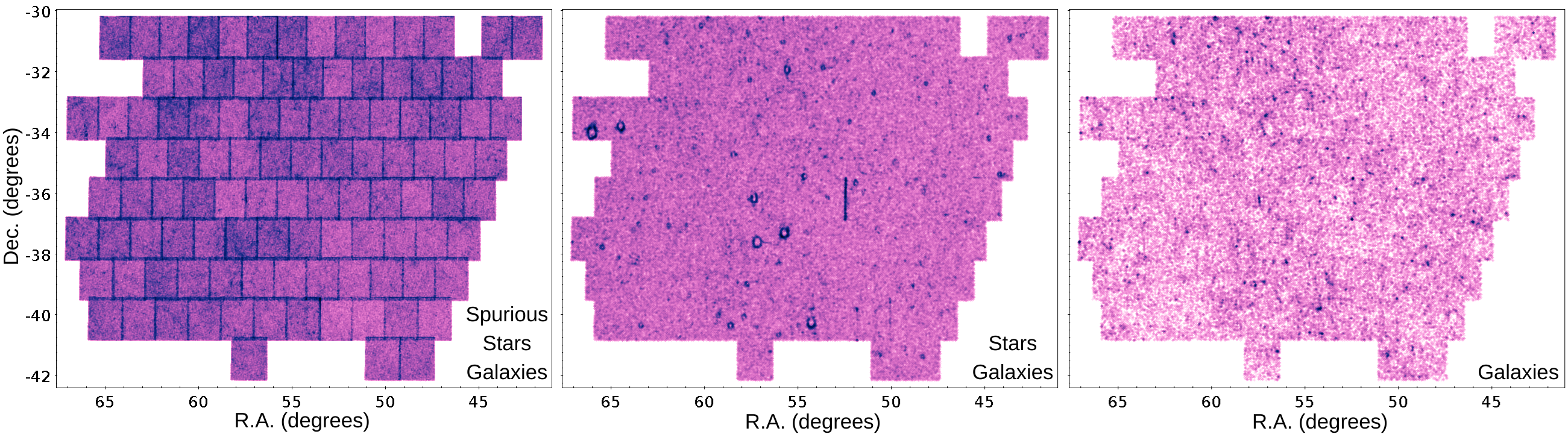}
    \caption{From left to right, the panels show the evolution of the spatial distribution of objects, from the catalog by \cite{Haack2024} {\it (left panel)} to the final S+FP extragalactic catalog {\it (right panel)} (see text for details).}
    \label{fig:proceso}
\end{figure*}

In Figure\,\ref{fig:proceso}, we show the evolution in the process of spurious cleaning, star/galaxy separation and flagging from the original photometric catalog. Specifically, in the left panel it can be seen the projected spatial distribution of the objects in the initial catalog of 3,000,000 (spurious, stars and galaxies) sources. In the central panel of Figure\,\ref{fig:proceso}, we show the spatial distribution of the objects in that catalog after cleaning from spurious sources. Finally, in the right panel, the spatial distribution of the final extragalactic catalog containing 119,580 galaxies is presented. It is noticeable how easy it is to visually identify in this last panel, large-scale substructures not seen in the other panels. 

To arrive at this final spatial distribution of 119,580 galaxies, 15,088 galaxies were flagged and separated. Upon reviewing the spatial distribution with 134,593 galaxies, abnormal overdensities were observed, and they correspond to sources with a significantly high background measurement. This occurs due to specific problems in the reduction process in one of the 106 S-PLUS pointings considered in this work. Furthermore, these overdensities appear in the peripheries of extremely bright stars, which are not well represented within the models utilized for the classification of spurious objects, stars, and galaxies. For details on this peculiar problem, see Appendix \ref{app:back_probl}.

\begin{figure}[h!]
    \centering
    \includegraphics[width=1.0\columnwidth]{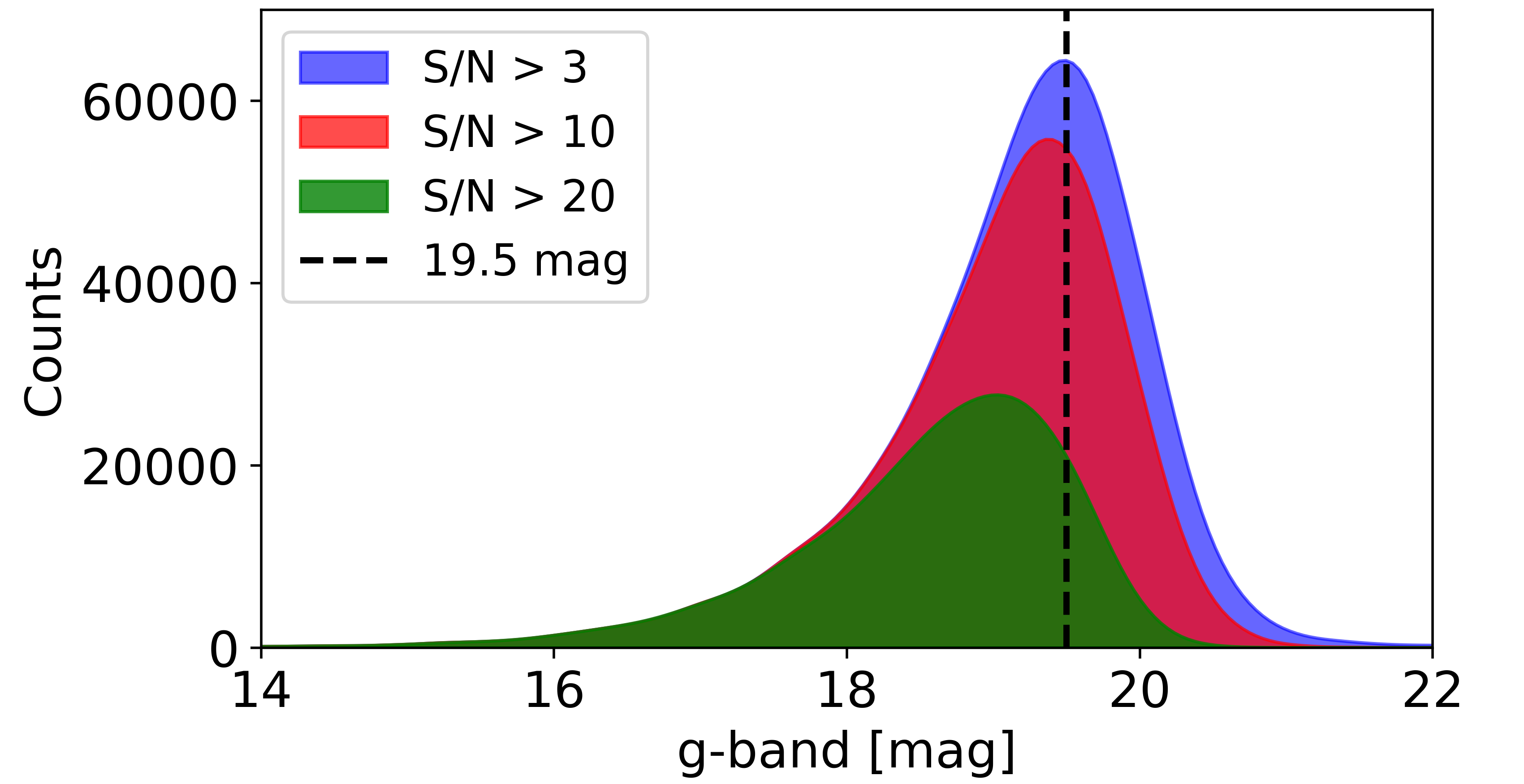}
    \caption{Histogram of g-band AUTO magnitudes and the choice of   photometric depth at 19.5 mag. The distributions correspond to three S/N thresholds (S/N > 3, blue; S/N > 10, red; and S/N > 20, green).}
    \label{fig:histo_r_photoz}
\end{figure}

Figure\,\ref{fig:histo_r_photoz} shows g-band AUTO magnitudes for the extragalactic catalog of 119,580 galaxies. It can be seen that the number of galaxies increases towards the faint end, with very few of them being detected after the peak at 19.5 mag. The black dashed vertical line corresponds to 19.5 mag in the g-band and sets the limit that we will take as photometric depth. 

\subsection{Estimation of background galaxy counts and completeness analysis}

A completeness analysis is fundamental to assess observational limitations of extragalactic catalogs and to ensure their representative galaxy sampling in cosmological studies. This requires an accurate estimation of background galaxy densities, particularly for wide-field surveys like S-PLUS. Our analysis focuses on a catalog covering a region of $208\ \text{deg}^2$ and with an apparent magnitude limit of $g \leq 19.5$ mag, due to the photometric depth in the g-band. In addition, we consider the redshift interval $0.01 \leq z_{\mathrm{spec}} \leq 1.0$. The lower limit of $z_{\mathrm{spec}} = 0.01$ explicitly excludes the Fornax cluster  ($z_{\mathrm{spec}} \approx 0.0047$) considering the radial velocity dispersion constraint given by \cite{Maddox2019}. The upper limit of $z_{\mathrm{spec}}=1.0$ reflects the detection threshold of S-PLUS, following \cite{Herpich2024}. In that sense, we isolate background populations, including field, group and cluster galaxies, along the line of sight beyond Fornax \citep{Blanton2001}. It is remarkable that our extragalactic catalog includes only 403 FLS galaxies, which are negligible (0.33$\%$) with respect to the 119,580 sources in the final sample.

To evaluate the completeness of our catalog, we used a simulated catalog from a synthetic sky light cone developed by \cite{Araya-Araya2021} to emulate S-PLUS data. The mock lightcones employed were constructed utilizing the \textsc{L-GALAXIES} semi-analytic model (SAM) as presented by \citet{Henriques2015}. This model is applied to the dark matter-only Millennium simulation \citep{Springel2005} to generate synthetic galaxies. The SAM operates on the merger trees from the Millennium simulation, which are constructed using the \textsc{SUBFIND} algorithm \citep{Springel2001}, ensuring that the simulated galaxy formation and evolution processes are grounded in a realistic dark matter distribution. To align with modern cosmological constraints, the model output is scaled to the \citet{Planck2014}'s cosmological parameters using the method described by \citet{Angulo2010}. The \textsc{L-GALAXIES} code incorporates a comprehensive set of astrophysical processes critical to galaxy evolution, such as gas infall, radiative cooling, star formation, metal enrichment, the growth of supermassive black holes, and feedback mechanisms from both supernovae and active galactic nuclei (AGN). For a complete description of these physical processes, we refer the reader to the Supplementary Material of \citet{Henriques2015}. The final output of the SAM provides essential physical properties for the synthetic galaxies, including stellar mass, gas mass, and SFR.

Under the aforementioned considerations regarding the projected area, photometric depth and redshift range adopted, the completeness (C) of our catalog is determined as follows:

\begin{equation}
C = \frac{N}{N_{\text{mock}}} = \frac{72,823}{101,017} \sim 72\%, 
    \end{equation}

\noindent where $N_{\text{mock}}$ is the number of galaxies expected according to the simulated catalog, and $N$ corresponds to the number of galaxies present in our catalog with g < 19.5 mag (which represents 60$\%$ of the entire catalog). Given cosmic variance, this value of completeness is reassuring.

\section{Properties of the S+FP Extragalactic Catalog}
\label{sec:properties}

In this section, we estimate $z_{\mathrm{phot}}$, stellar masses, SFRs and the $D4000_{N}$ index values by using a ML approach implementing Random Forest regression \citep{Breiman2001}, an ensemble method that combines multiple decision trees to enhance predictive stability. This algorithm, particularly effective for high-dimensional photometric problems \citep{CarrascoKind2014}, builds independent trees where each node splits the feature space by minimizing the mean squared error (MSE) of the estimated variable. The final prediction results from averaging the individual estimates of 500 trees, controlling complexity with a maximum depth of 20 and a minimum of 5 samples per split to prevent overfitting. The models were trained (independently for each parameter) over the 22 AUTO magnitudes, corresponding to the S-PLUS filters combined with GALEX, VHS and WISE, added to the 231 photometric colors that arise from this combination.

The preprocessing systematically addressed challenges inherent to multi-survey data. First, we removed features with more than a 30\% of missing values, preserving those with sufficient coverage to ensure representativity. Subsequently, missing values were imputed using feature-wise medians, which are robust against outliers, and a standard scaling (mean=0 and standard deviation=1) was applied to normalize the distributions. This pipeline ensured that intrinsic differences in photometric scales between surveys did not bias the model towards brighter bands.

\subsection{Photometric redshfits}
\label{sec:zphot}

In the context of the study of the Fornax cluster   and the large-scale structure in its surroundings, it is necessary to obtain our own estimates of $z_{\mathrm{phot}}$. This is because the redshift of Fornax itself ($z_{\mathrm{spec}}$ = 0.0046) is smaller than the errors in the estimates of $z_{\mathrm{phot}}$ reported in the literature. To assess the quality of the $z_{\mathrm{phot}}$ estimates, the calculation of the Normalized Median Absolute Deviation ($\sigma_{NMAD}$) of the bias $\Delta$z = $z_{\mathrm{phot}}$ - $z_{\mathrm{spec}}$ is commonly used. Following \citet{Brammer2008}, $\sigma_{NMAD}$ is defined as:
\begin{equation}
\label{eq:sigmanmad}
    \sigma_{\text{NMAD}} = 1.48 \times \text{median} \left( \frac{\Delta z - \text{median}(\Delta z)}{1 + z_{\text{spec}}} \right)
\end{equation}
It is worth mentioning that, different to the standard definition of $\sigma_{NMAD}$ given by \citet{Ilbert2006} and \citet{Li2022}, Eq. \eqref{eq:sigmanmad} is less sensitive to outliers (also known as catastrophic errors) that, according to \cite{Ilbert2006}, are galaxies that satisfy:
\begin{equation}
\label{eq:outliers}
\eta = \frac{|\Delta z|}{1 + z_{\text{spec}}} > 0.15
\end{equation}

Just to give some reference values, \cite{Hernan-Caballero2021} use SED-Fitting and 60 photometric bands from MiniJPAS \citep{Bonoli2021} to obtain a $\sigma_{NMAD}$ $\sim$ 0.013.  Using a ML approach and fewer photometric bands, \cite{Lima2022} achieves a $\sigma_{NMAD}$ $\sim$ 0.023 with the S-PLUS survey. Based on a DL approach, \citet{Teixeira2024} achieved a $\sigma_{NMAD}$ $\sim$ 0.0293 with the DECam Local Volume Exploration (DELVE) Survey. Therefore, regardless of the approach, this means that for the specific case of Fornax, $z_{\mathrm{phot}}$ estimates cannot be trusted since their errors are larger than the cluster redshift itself. In this sense, our goal will be to find a lower limit ($z_{\mathrm{lim}}$) for our $z_{\mathrm{phot}}$ estimates. That will allow us to clean up our extragalactic catalog from background galaxies rather than selecting cluster member candidates. The use of our own model instead of the one provided by \cite{Lima2022} is justified since not all galaxies present in our catalog are detected or characterized in that work, as already explained in Section\,\ref{sec:data}.

In order to achieve a good estimation of such a limit, we have complemented the 12 S-PLUS bands with information at both UV and IR wavelengths. We have used the public catalogs of GALEX, VHS-VISTA and AllWISE. The combination of filters used and their respective transmission curves are shown in Appendix \ref{app:filter}. The magnitudes were transformed to the standard AB system \citep{Oke1974} and were corrected for extinction E(B-V) following \citet{Schlafly11}. 

\begin{figure}[h!]
    \centering
    \includegraphics[width=0.9\columnwidth]{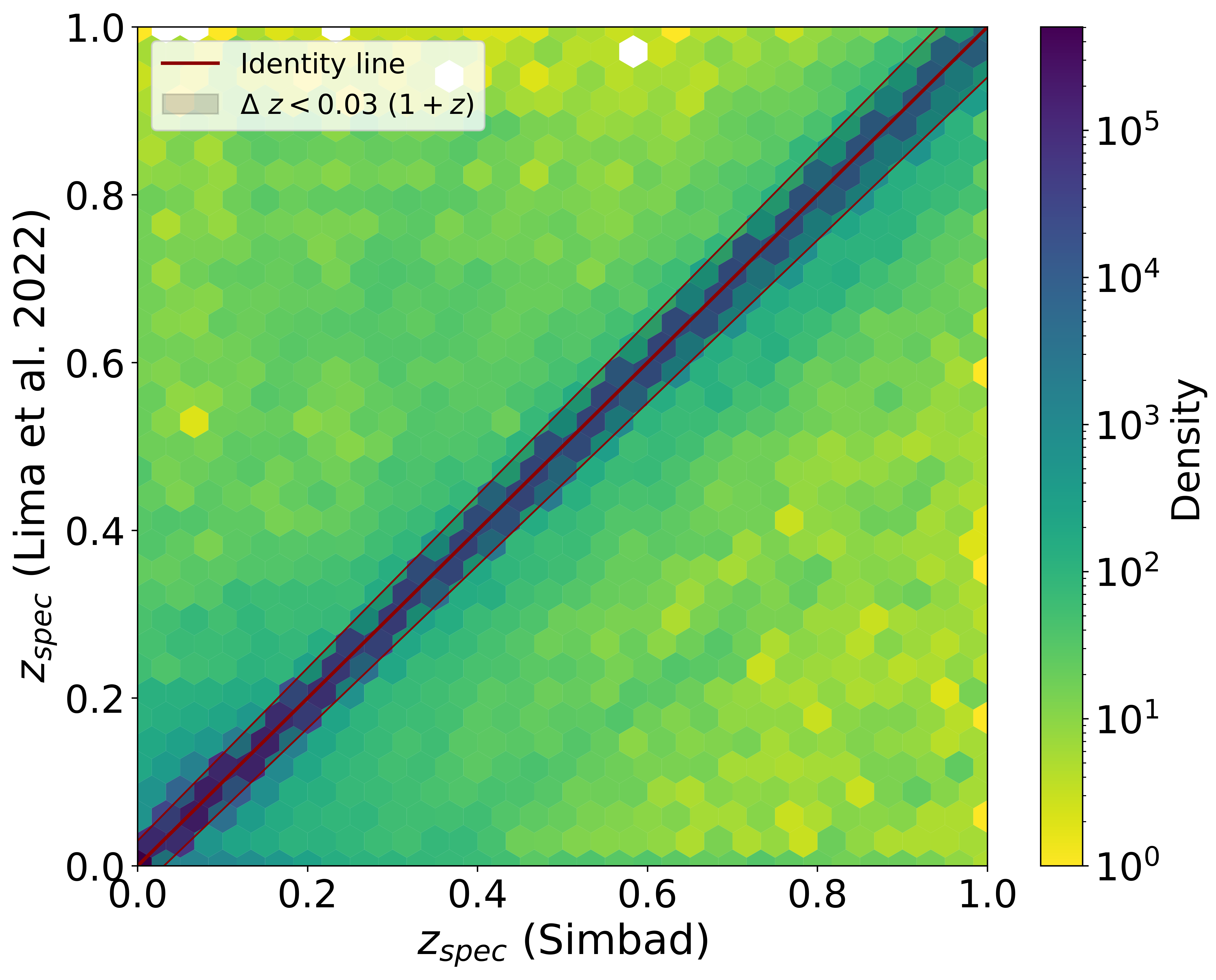}
    \caption{Comparison of $z_{\mathrm{spec}}$, provided by SIMBAD and \cite{Lima2022}, that were utilized to build the spectroscopic sample used to validate the $z_{\mathrm{phot}}$ estimates. The red line is the identity line. The region colored in gray corresponds to the sources that display $\Delta z_{\mathrm{spec}} < 0.03$. The galaxies in that region constitute a double-checked sample that includes $85\%$ of the total sources in the plot.}
    \label{fig:zspec comparison}
\end{figure}

To validate our estimations, we selected galaxies included in our extragalactic catalog that have spectroscopic radial velocities. For such a selection, we performed a crossmatch between SIMBAD and an all-sky spectroscopic compilation given by \citet{Lima2022}. The idea of performing such a crossmatch is to obtain more confidence in the spectroscopic values used for the validation of our results. As it can be seen in Figure\,\ref{fig:zspec comparison}, there is a certain amount (5$\%$) of spectroscopic redshifts that display disagreements in the values provided by different sources. 
Therefore, we have only considered for training and validation the galaxies located in the region marked in gray in Figure\,\ref{fig:zspec comparison}, that is, those displaying a dispersion of $\pm$ 0.03 from the identity line. 

\begin{figure}[h!]
    \centering
    \includegraphics[width=1\columnwidth]{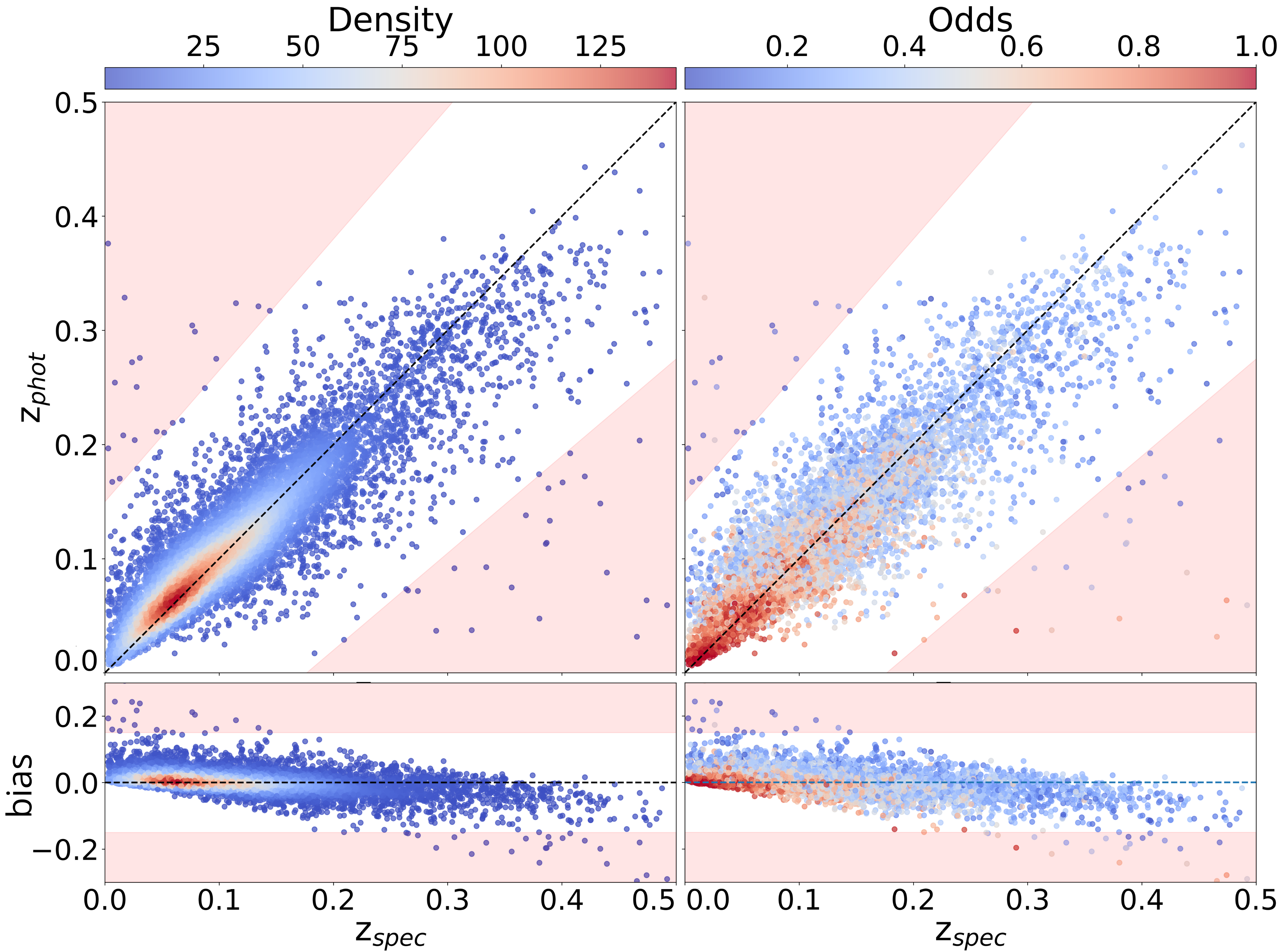}
    \caption{The left panel shows the comparison between $z_{\mathrm{spec}}$ and $z_{\mathrm{phot}}$ obtained with a ML approach colored by source density. In the right panel, the comparison is colored by {\it Odds}. Below each panel, we also show the corresponding bias. The regions colored in red correspond to the outliers region, according to Eq. \eqref{eq:outliers}. The black lines correspond to the identity lines.}
    \label{fig:photozs_machine_odds}
\end{figure}

After training on 80\% of the galaxies ($z \leq 0.5$) and validating on the remaining 20\%, the model achieved $\sigma_{\mathrm{NMAD}} = 0.0214$, $\eta = 0.42\%$ and a bias of 0.0025 (Figure \ref{fig:photozs_machine_odds}). The bias subplot evidences a minor overestimation for galaxies at lower redshifts and an underestimation for those at higher redshifts. This trend is also found in \cite{Lima2022}.

To quantify uncertainties, we implemented two complementary metrics: $\sigma_{68}$, the standard deviation that encompasses 68\% of the ensemble predictions, and \textit{Odds}, defined as the integrated probability density function (PDF) within $z_{\mathrm{phot}} \pm 0.02$. Formally, 

\begin{equation}
\mathrm{Odds} = \int_{z_{\mathrm{phot}} - 0.02}^{z_{\mathrm{phot}} + 0.02} \mathrm{PDF}(z) \, dz.
\end{equation}

\begin{figure}[h!]
    \centering
    \includegraphics[width=1\columnwidth]{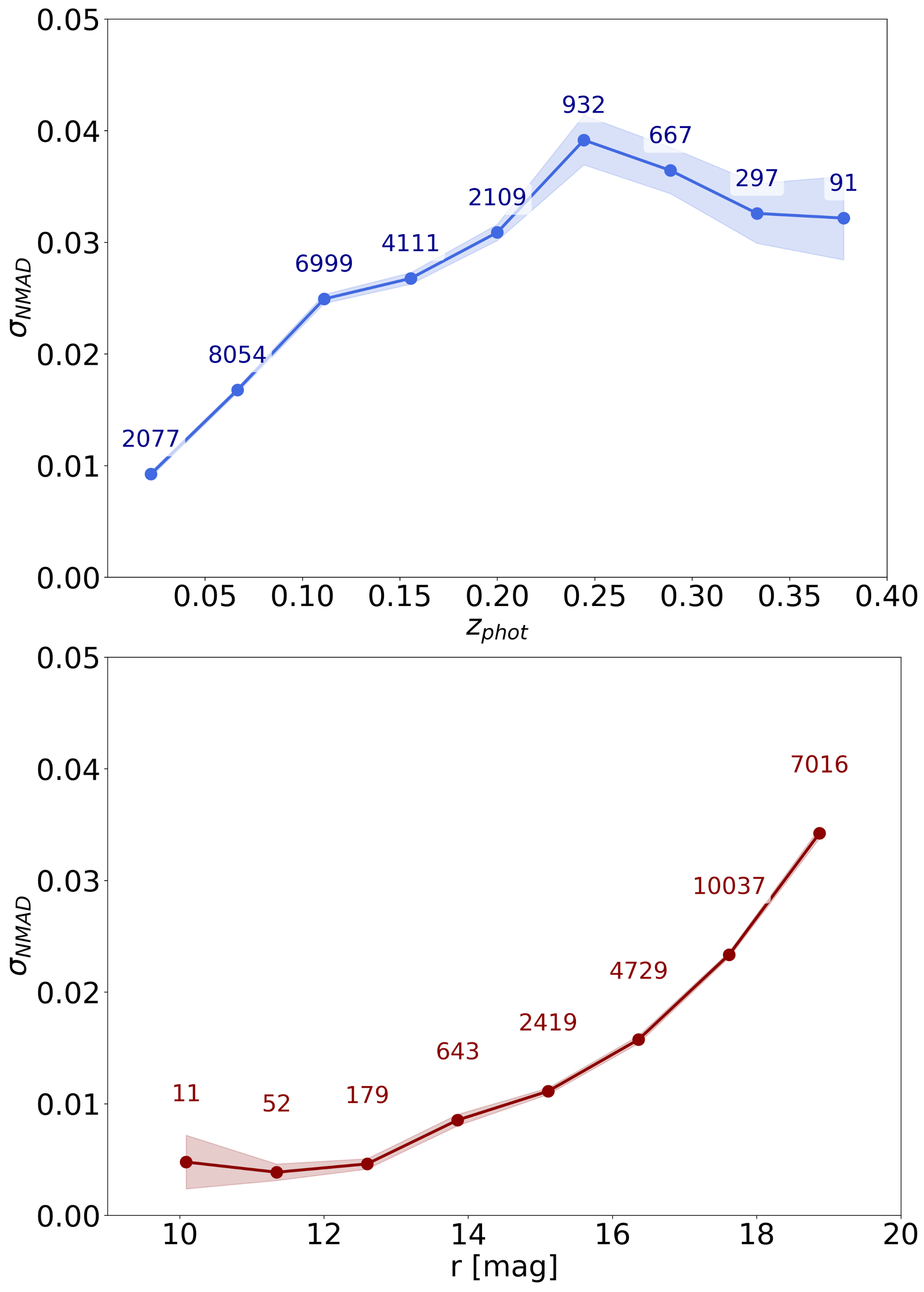}
    \caption{The top panel presents the evolution of $\sigma_{NMAD}$ with $z_{\mathrm{phot}}$ at different $z_{\mathrm{phot}}$ bins. Above each bin we show the total number of galaxies considered in the calculation. Error bars and interpolation were calculated with bootstrapping. In the lower panel, we show the variation of $\sigma_{NMAD}$ with $r_{auto}$.}
    \label{fig:sigmaNMAD_evolucion}
\end{figure}

While $\sigma_{68}$ maps the absolute width of the probability distribution, \textit{Odds} quantifies its relative concentration near the peak: values close to 1 indicate narrow PDFs and a high confidence. In Figure\,\ref{fig:photozs_machine_odds}, we observe that galaxies with high \textit{Odds} ($>0.8$) predominantly cluster at $z < 0.2$, closely following the perfect-fit relation. In contrast, at $z > 0.3$, \textit{Odds} systematically decreases ($<0.6$) and the PDFs broaden, reflecting the increased difficulty in discriminating spectral features in distant galaxies with faint fluxes. This behavior correlates with the increase of $\sigma_{\mathrm{NMAD}}$ as a function of redshift (top panel of Figure\,\ref{fig:sigmaNMAD_evolucion}). There, the dispersion grows from 0.01 ($z<0.01$) to 0.025 ($z=0.1$), reaching a peak at $z=0.25$, and then dropping to 0.03 near $z=0.4$. The error of $\sigma_{\mathrm{NMAD}}$ in each bin is estimated by bootstrapping \citep{Efron1979}, i.e., recalculating the estimator over multiple random resamplings with replacement of the residuals, and taking the resulting dispersion as uncertainty.

The bottom panel of Figure\,\ref{fig:sigmaNMAD_evolucion} shows that $\sigma_{\mathrm{NMAD}}$ remains stable ($\sim0.01$) up to $r \sim 14.0$ mag, after which it increases sharply to $\sim0.035$ at $r\sim19.5$ mag. This critical transition appears because, for fainter galaxies, the photometric errors grow degrading the model's ability to discern subtle variations in optical-IR colors. The correlation between \textit{Odds}, r-band magnitudes, and $z_{\mathrm{phot}}$ is detailed in the Appendix\,\ref{app:odds}. 

Given these results, we set $z_{\mathrm{lim}}$ $\sim$ 0.03 as a lower limit to separate background galaxies and Fornax cluster  candidates, adopting a 3$\sigma$ criterion on the $\sigma_{\mathrm{NMAD}}$ $\sim$ 0.01 value ($z < 0.01$) already mentioned. In that sense, we have been able to find 350 new Fornax member candidates, all of them without measured $z_{\mathrm{spec}}$ and that are not included in the FLS. 

\subsection{Stellar masses}
\label{sec:masses}
Robust stellar mass estimates are critical for understanding galaxy formation and evolution across diverse environments. Stellar mass serves as a fundamental parameter that links observed galaxy properties, such as SFRs, metallicities, and morphologies, to their underlying physical processes and dark matter halo assembly histories (e.g., \citealt{Behroozi2019}). In dense environments such as clusters, accurate masses allow the study of quenching mechanisms (e.g. \citealt{Peng2010}), while in the field they help isolate secular evolutionary pathways. Furthermore, stellar mass functions in different environments constrain hierarchical structure formation models (e.g. \citealt{Wechsler2018}), although systematic errors in mass estimation can bias comparisons between observations and simulations \citep{Leja2019}. Thus reliable mass determinations are essential for probing environmental dependencies, galaxy-halo connections, and the role of feedback in shaping the galaxy population.

Here, stellar masses have been estimated considering the same ML architecture presented in Section\,\ref{sec:zphot}. To validate the estimations, we have again considered an S-PLUS catalog of the Stripe-82 region with the same characteristics as the catalog in the Fornax direction. We took advantage of the information given by SDSS DR8, in particular by the catalogs of galaxy properties provided by the MPA-JHU group, described in \cite{Kauffmann2003}, \cite{Brinchmann2004} and \cite{Tremonti2004}.
From them, we chose ‘$lgm\_tot\_p50$’, the median estimate of the logarithm of the total stellar mass PDF, and matched it with the magnitudes, errors and colors of the multiple photometric bands used in this work. To these learning features we now add the redshift. The range of possible stellar masses for training has been limited to 8.5 - 12.5 log$(M/M_\odot)$.

\begin{figure*}[h!]
    \centering
    \includegraphics[width=0.85\textwidth]{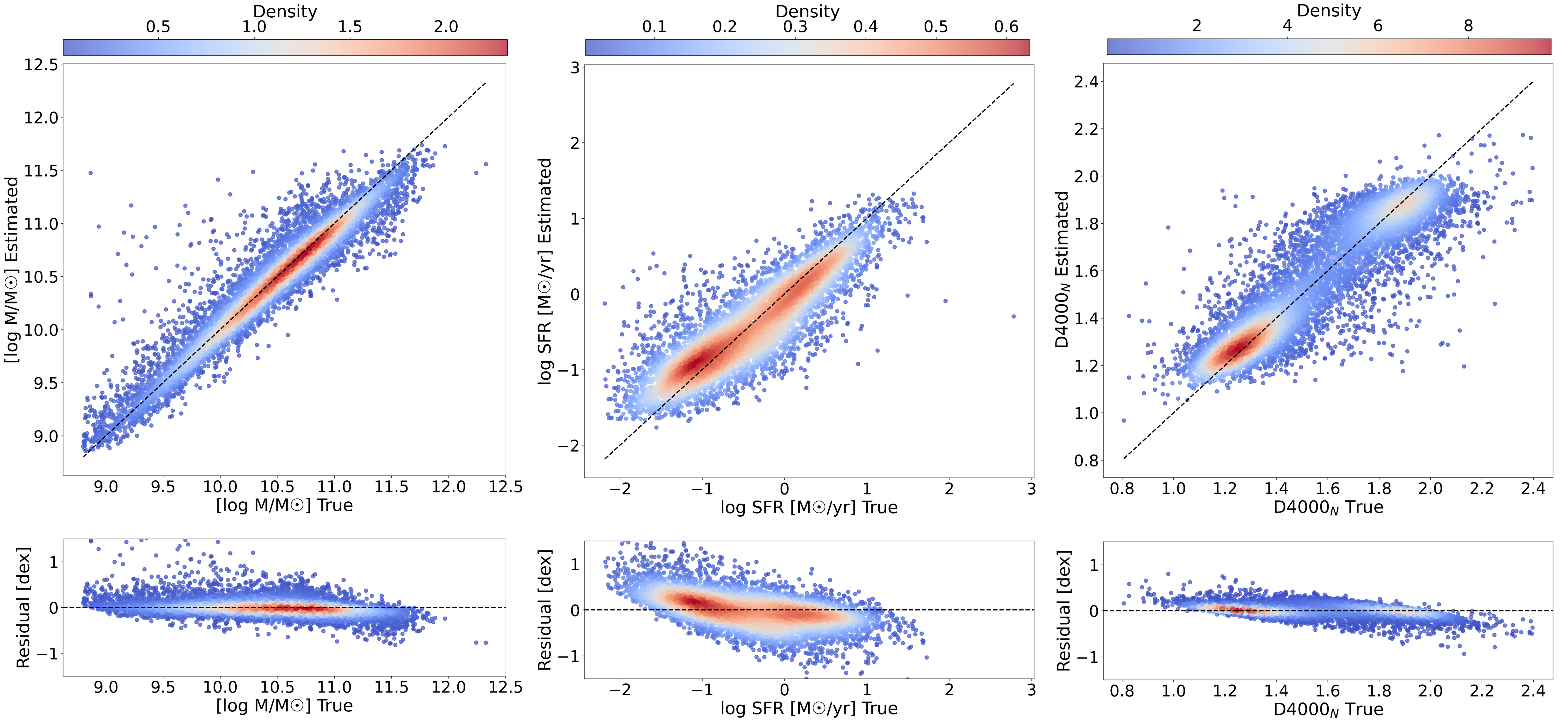}
    \caption{Comparison between the true values provided spectroscopically by SDSS DR8, and the values predicted using ML, for stellar mass {\it (left)}, SFR {\it (center)} and $D4000_{N}$ index {\it (right)}. Each plot is colored by density and the black lines correspond to the identity lines. Each lower subplot represents the residual of the respective estimated property.}
    \label{fig:masasML}
\end{figure*}

The stellar masses obtained are shown in the left panel of Figure\,\ref{fig:masasML}, evidencing a good estimation, with a determination coefficient ($R^2$) of 89$\%$. $R^2$ is a metric that indicates the proportion of variance in the dependent variable that is explained by the regression model. It is worth noticing that there is no significant bias in the whole mass regime considered in our analysis. Due to its precision and the applicability over the whole sample (that is, the model keeps its precision regardless of the quality of the photometry), we will take the estimates of this approach to characterize our catalog.

\subsection{SFR and $D4000_{N}$ index}
\label{sec:SFR-D4000}
Aiming at performing a separation between quiescent and star-forming galaxies in the extragalactic catalog, we estimated SFRs and $D4000_{N}$ index values (an indicator of the mean stellar age and metallicity of a galaxy) with a ML approach using the same architecture already explained.

For the SFR estimation, and from the galSpecExtra SDSS DR8 catalog, we used the parameter ‘$sfr\_tot\_p50$’, that is, the median estimate of the logarithm of the total SFR PDF. This parameter is derived by combining emission line measurements within the spectroscopic fiber of SDSS, where possible, and considering aperture corrections following \citet{Gallazzi2005} and \citet{Salim2007}. For those objects where the emission line flux within the fiber do not provide an estimate of the SFR, model fits to the integrated photometry were performed, learning the redshift and ‘$lgm\_tot\_p50$’ from the photometric features for a total of 32,720 galaxies. The range of possible estimates was limited to $-2.2 < log(SFR~ [M_\odot/yr]) < 3.0 $. The estimates are shown in the middle panel of Figure\,\ref{fig:masasML}, achieving a $R^2$ of 71$\%$ but displaying a slight overestimation for galaxies with $log(SFR~ [M_\odot/yr]) < -1$.

For the estimation of the $D4000_{N}$ index, and from the galSpecIndx SDSS DR8 catalog, we used the parameter ‘$D4000_{N}$’ as defined by \cite{Balogh1999}. From the photometric features, we learned the redshift and the ‘$lgm\_tot\_p50$’ and '$sfr\_tot\_p50$' parameters for a total of 46,235 galaxies. In this way, a sequential and nested learning was performed, that is, features are added for the next learning, following the methodology presented by \cite{Euclid2025}. Here we limited the possible estimation range to $0.5 < D4000_{N} < 3.0$. The estimates are shown in the right panel of Figure\,\ref{fig:masasML}, achieving a $R^2$ of 81$\%$.

\begin{figure*}[h!]
    \centering
    \includegraphics[width=0.85\linewidth]{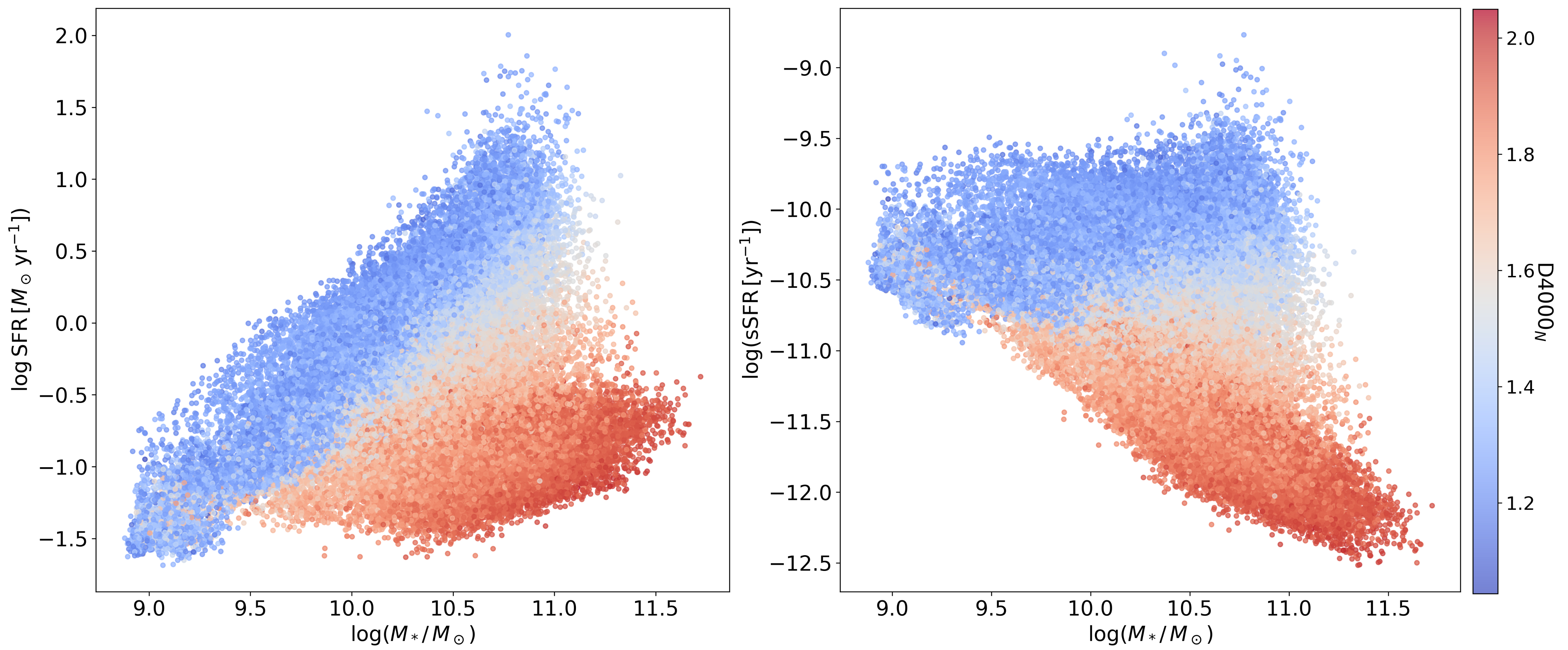}
    \caption{Stellar mass-SFR relation showing the SFMS {\it (left)} and stellar mass-sSFR relation {\it (right)}. Both plots share the same color bar, representing $D4000_{N}$ index values.}
    \label{fig:quiescent_sf}
\end{figure*}

Considering our stellar mass, SFR and $D4000_{N}$ estimates, in the left panel of Figure\,\ref{fig:quiescent_sf} we show the stellar mass-SFR relation and, in the right panel, the stellar mass-specific SFR (sSFR) relation. The left panel confirms the established correlation where star-forming galaxies (SFGs) follow a tight sequence spanning the intervals of $\log(M/M_\odot) < 11$ and $-1.0 < \log(\text{SFR}~[M_\odot/\text{yr}]) < 1.5$, which is commonly known as Star-Forming Main Sequence (SFMS). Quiescent galaxies deviate for $log⁡(SFR~ [M_\odot/yr]) < −1.0$ \citep{Noeske2007,Speagle2014}. The right panel reveals a clear bimodality in sSFR: SFGs are found at $log⁡(sSFR [yr^{-1}]) > -10.5$, contrasting with quiescent systems found at $log⁡(sSFR[yr^{-1}]) < -11.0$ \citep{Peng2010}. It can be seen that $D4000_{N}$ values robustly separate these regimes: SFGs show young populations ($D4000_{N} < 1.5$; \citealt{Kauffmann2003}), while quiescent galaxies exhibit evolved stellar components ($D4000_{N} >1.6$; \citealt{Ilbert2013}). The transitional {\it green valley} ($1.5 \leq D4000_{N} \leq 1.6$) suggests ongoing quenching, likely driven by mass-dependent processes such as gas depletion \citep{Leroy2008} or environmental effects \citep{Peng2015}. Adopting these criteria, in our catalog (119,580 galaxies) we have identified 51,390 ($43\%$) quiescent galaxies, 46,586 ($39\%$) star-forming galaxies and 21,604 ($18\%$) transition galaxies. This integral fraction of transitional systems is consistent with the $\sim$15$\%$ reported for the Local Universe \citep{Schawinski14} and the 15–20$\%$ typical of mass-complete samples at intermediate redshifts \citep{Muzzin13, Tomczak14}.

These results are consistent with the idea that stellar mass is the main regulator of galaxy evolution at low redshifts ($z \leq 0.5$), with the SFMS defining the evolutionary pathway for star-forming systems. The $D4000_{N}$ bimodality confirms its efficacy as an age proxy in low-z surveys. Future spatially resolved $D4000_{N}$ mapping could disentangle quenching mechanisms (for example, AGN feedback versus environmental stripping; \citealt{Bluck2016}).

\subsection{Emission line galaxies}
Identifying ELGs is crucial as they serve as direct tracers of ongoing star formation and nuclear activity, providing key insights into galaxy evolution and serving as efficient probes of large-scale structure. The S-PLUS J0660 filter captures the H${\alpha}$+[NII] lines at the distance of Fornax. Consequently, an excess in this filter indicates emission. However, it can also contain other lines, such as [OIII] (5007\,\r{A}), for galaxies at $z_{spec} \sim 0.32$. Without redshift information, though, it is not possible to determine which lines are responsible for an observed excess.

Following \citet[][hereafter G25]{Gutierrez-Soto2025}, we identify 181 of such galaxies in the S+FP region. This method was originally developed and optimized to detect high flux excess for point sources, but we have successfully applied it to extended objects and detected  a J0660 excess in galaxies. Comparing with the 77 ELGs presented by \citet[][hereafter L25]{Lopes2025} only 16 FLS galaxies are shared between both methods. This is because G25 fit the stellar locus in the $(r - J0660)$ versus $(r - i)$ color–color diagram. The scatter~$\sigma$ for each source is estimated via an approximation to the full error‐propagation formalism. We then apply a 3$\sigma$ threshold, and flag as an ELG candidate any object lying more than 3$\sigma$ above the fitted locus. The choice of 3$\sigma$ follows a standard convention to minimize false positives. In contrast, L25 applies the Three Filter Method (e.g. \citealt{Vilella-Rojo2015}) to S-PLUS images in order to create H${\alpha}$+[NII] emission line maps of 77 spectroscopically confirmed galaxy members of Fornax. This technique detects a much wider range of $(r-J0660)$ color, including the identification of objects with moderate intensity of emission. Therefore, galaxies with a moderate or extreme excess are simultaneously selected as ELGs in this case. 

\begin{figure}[h!]
    \centering
    \includegraphics[width=1\columnwidth]{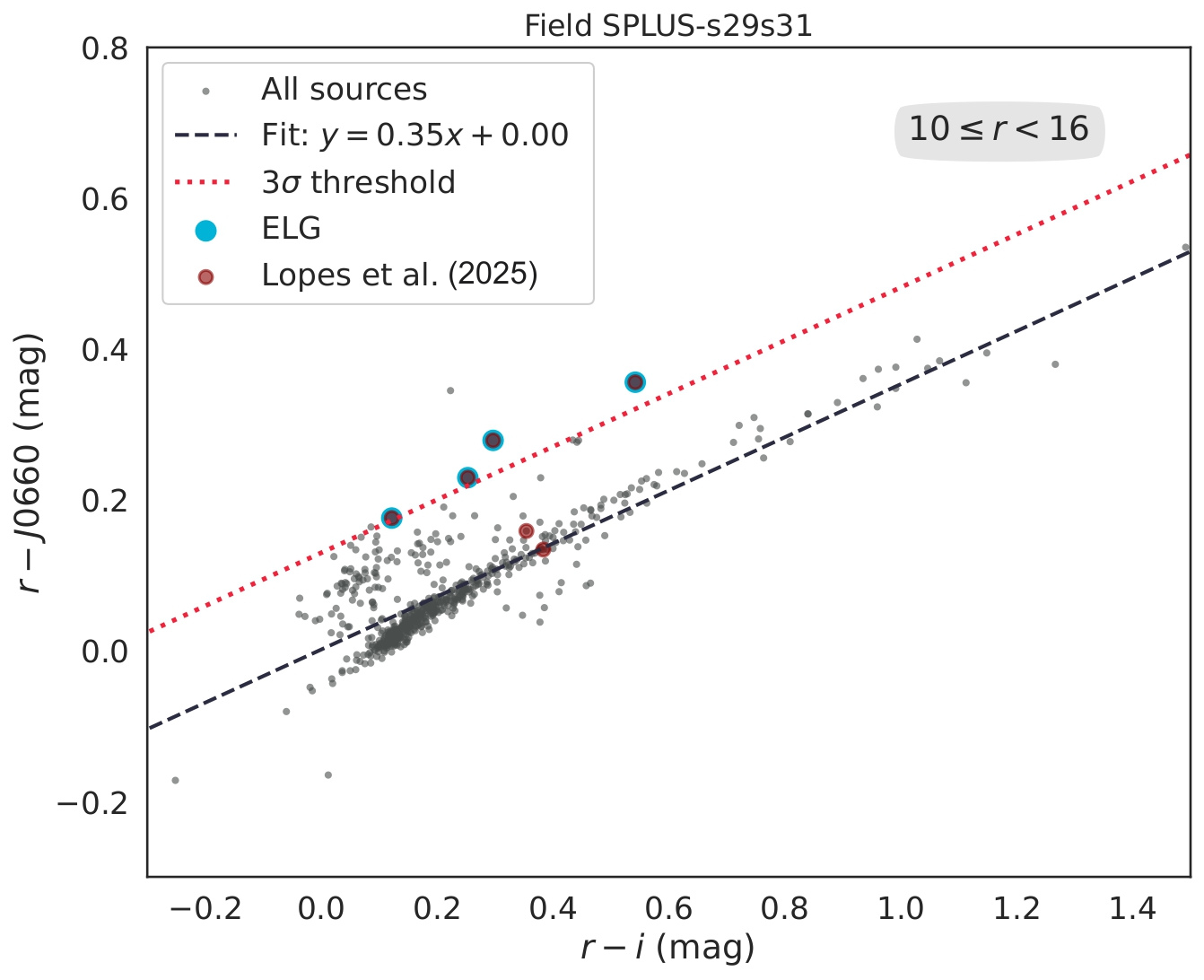}
    \caption{The ($r-J0660$) versus ($r-i$) color-color diagram for the field ID SPLUS-s29s31 with $10~\text{mag} < r-band < 16 ~\text{mag}$, considering all objects (stars and galaxies). The black dashed line corresponds to the stellar locus fit and the red dotted line represents a 3$\sigma$ deviation from the stellar locus. We depict in cyan 4 ELGs identified by G25 and, in red, 6 ELGs by L25.}
    \label{fig:ELG}
\end{figure}

Figure\,\ref{fig:ELG} shows an example of ELG identification by the method of G25, with candidates highlighted in cyan for a specific field (SPLUS-s29s31) and in the limited magnitude range of $10\,\mathrm{mag}<r\text{-band}<16\,\mathrm{mag}$. The ELGs detected by L25 are highlighted in red in the plot. It should be taken into account that the sample of L25 includes only Fornax member galaxies and that the method of G25 was applied on the whole extragalactic catalog without taking into account, in many cases, the radial velocity of the galaxy. Considering that both methods seem to be compatible for galaxies with extreme excess, G25 method recovers completely what has already been identified by L25. This allows us to quickly identify targets for spectroscopic follow up, without knowing the radial velocity of the galaxies. It is necessary to comment that, ignoring the redshift, the excess flux may correspond to other possible emission lines, not necessarily to H${\alpha}$+[NII]. It is worth noting that in Figure\,\ref{fig:ELG} there are sources above the red dotted line, which is an average representation of the individual $\sigma$, that are not marked as ELGs. This is because these sources are stars, not galaxies. All stars in the field were used to compute the locus fit (solid black line), so they are displayed in the plot.

\subsection{eROSITA X--ray clusters identification}

We use the sample of galaxy clusters and groups from eROSITA All-Sky Survey Data Release 1 (hereafter eRASS1) from \cite{Bulbul2024} and \cite{Kluge2024}, in combination with the X--ray morphological information from \cite{Sanders2025}, to identify clusters in the 208 deg$^2$ area in the direction of the Fornax cluster. The cluster and group catalog has known statistical contamination (estimated purity of the sample $\sim 86$\%) due to the shallow depth of the survey and false identification of extended X-ray sources. Most of these contaminants have low richness and low or high redshifts. Therefore, to remove contaminants, we add additional constraints to the sample. We limit the identification by applying a criterium similar to those used by \cite{Zenteno2025}: {\it (i)} clusters located inside a region of $43^{\circ}< \mathrm{RA\_XFIT} < 67^{\circ}$ and $-42^{\circ} < \mathrm{DEC\_XFIT} < -30^{\circ}$; {\it (ii)} redshift between $0.05 \le \mathrm{BEST\_Z} \le 0.4$; {\it (iii)} photometric redshifts smaller than the local limiting redshift (IN\_ZVLIM=True); {\it (iv)} normalize richness LAMBDA\_NORM $\geq15$; {\it (v)} probability of the cluster being a contaminant PCONT $< 0.1$;  {\it (vi)} fraction of the cluster area masked MASKFRAC $< 0.1$;  {\it (vii)} cluster mass $M_{500} \ge 5 \times 10^{13}$ M$_\odot$; and {\it (viii)} $R_{500} > 0$. After applying the selection criteria, the final sample contains 158 clusters within the above area centered on NGC 1399.

\begin{figure*}[h!]
    \centering
    \includegraphics[width=1.0\linewidth]{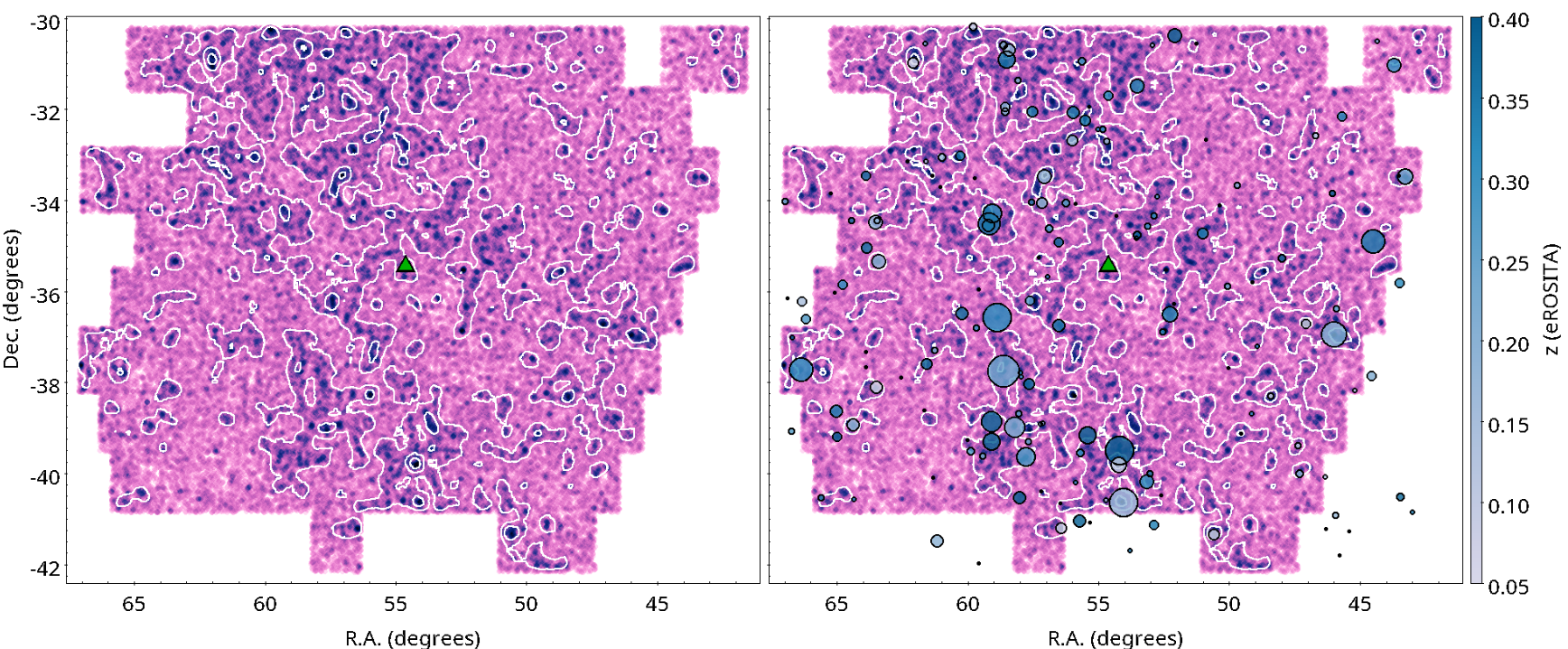}
    \caption{Projected overdensities of the extragalactic catalog {\it (left)} superimposed with galaxy clusters identified by eROSITA marked as circles {\it (right)}. The sizes of the circles are proportional to $M_{500}$ and the colors correspond to the BEST\_Z redshift provided by eRASS1. The central green triangle indicates the position of NGC\,1399.}
    \label{fig:eROS}
\end{figure*}

The left panel of Figure\,\ref{fig:eROS} shows the surface density of galaxies in our catalog with white isodensity contours highlighting the projected overdensities. In the right panel, circles at the positions of the eROSITA clusters are superimposed on the spatial distribution of the overdensities. The sizes of the circles are proportional to $M_{500}$ and the colors correspond to the BEST\_Z redshift provided by eRASS1 (color bar at the right of the plot). Notably, the projected spatial distribution of the eRASS1 clusters shows a significant degree of coincidence with the overdensities identified in our extragalactic catalog up to $z\sim0.4$. This strong spatial agreement provides a crucial validation of the robustness of our extragalactic catalog in tracing authentic large-scale structure. A subsequent analysis will go even deeper by combining these spatial matches with the galaxy properties from our catalog, such as stellar mass, SFR, and the D4000N index, to identify substructures within redshift bins and to classify the dynamical state of the sample using optical and X-ray morphological information. This will be complemented by 3D clustering algorithms (using R.A., Dec, and z) for a comprehensive investigation of galaxy evolution in these environments.

\section{Summary and conclusions}
\label{sec:conclusions}

We present an extragalactic catalog of 119,580 galaxies covering an area of 208 deg$^2$ in the direction of the Fornax cluster in 12 photometric bands, obtained through automatic learning algorithms. The NN models used have an accuracy and F1-score of 95$\%$ for the cleaning of spurious objects and star/galaxy separation. The format of the catalog is presented in Appendix\,\ref{sec:format}.

The completeness of the catalog has been estimated by comparison with a mock catalog, indicating 72 $\%$ of completeness with respect to the expected galaxies in the covered sky area, photometric g-band depth of 19.5 mag and 0.01< $z_{\mathrm{spec}}$ < 1.0.

From a ML approach, combining the 12 S-PLUS optical filters with data in the UV (GALEX) and IR (VHS and WISE), we have calculated $z_{\mathrm{phot}}$, reaching $\sigma_{\mathrm{NMAD}}\sim 0.02$ for the whole S+FP extragalactic catalog. For those galaxies with $z_{\mathrm{phot}}\sim 0.01$, $\sigma_{\mathrm{NMAD}}$ improves to 0.01. This allows us to set a lower limit of $z_{\mathrm{lim}}\sim 0.03$, under a 3$\sigma$ criterion, to separate Fornax candidate galaxies from background galaxies. In that sense, we have been able to find 350 new Fornax member candidates. In order to confirm them as real Fornax members, spectroscopic follow-ups or spectroscopic surveys like CHANCES  \citep{MendesHernandez2025}, will be needed. Additionaly, we provide $z_{\mathrm{phot}}$ for 119,230  background galaxies, including 8,226 with reported $z_{\mathrm{spec}}$.

The stellar mass values presented in the catalog correspond to those obtained from a ML approach, using spectroscopic catalogs from SDSS DR8.
The galaxy properties provided in the catalog have been extended by adding SFR and $D4000_{N}$ estimates. Together with the stellar masses, those values allowed us to analyze the SFMS, stellar mass-sSFR relation, and to classify the galaxies into quiescent, star-forming and transition objects. From these estimations, we have found that the S+FP extragalactic catalog contains 51,390 ($43\%$) quiescent galaxies, 46,586 ($39\%$) star-forming galaxies and 21,604 ($18\%$) transition galaxies. A total of 181 ELG candidates have been identified by the method presented in G25 to detect objects displaying a high excess flux in J0660.

In order to gain a deeper understanding of the large-scale structure around the Fornax cluster  and to identify the substructures that might be feeding it, in future works we plan to perform a detailed structure analysis by redshift bins taking advantage of the $z_{\mathrm{phot}}$ estimates presented here, together with the spatial distribution of physical properties such as stellar mass, SFR, age and metallicity. In that sense, it is worth noticing that in figure 14 of \cite{Lomeli2025} and in figure 7 of L25, overdensities of globular cluster candidates and ELGs, respectively, are evident in the north-west direction from the center of Fornax. Such an overdensity is also apparent in the spatial distribution of the S+FP extragalactic catalog, as it can be seen in the right panel of Figure\,\ref{fig:proceso}. In addition, we expect to extend this work to the complete Eridanus-Fornax-Doradus filament for which S-PLUS data have been already obtained.

\section{Data availability}
The data underlying this article will be shared on a reasonable request basis by the corresponding author.

\begin{acknowledgements}
RFH, AVSC, ARL, LAGS and JPC acknowledge financial support from Consejo Nacional de Investigaciones Científicas y Técnicas (CONICET), Agencia I+D+i (PICT 2019–03299) and Universidad Nacional de La Plata (Argentina). RFH thanks CAPES for financial support under the program Move La America 2025. AVSC thanks Fundação de Amparo à Pesquisa do Estado de São Paulo (FAPESP) for the support grant 2025/05085-1. R.D. gratefully acknowledges support by the ANID BASAL project FB210003. ERC acknowledges the support of the international Gemini Observatory, a program of NSF NOIRLab, which is managed by the Association of Universities for Research in Astronomy (AURA) under a cooperative agreement with the U.S. National Science Foundation, on behalf of the Gemini partnership of Argentina, Brazil, Canada, Chile, the Republic of Korea, and the United States of America.

The S-PLUS project, including the T80-South robotic telescope and the S-PLUS scientific survey, was founded as a partnership between the Fundação de Amparo à Pesquisa do Estado de São Paulo (FAPESP), the Observatório Nacional (ON), the Federal University of Sergipe (UFS), and the Federal University of Santa Catarina (UFSC), with important financial and practical contributions from other collaborating institutes in Brazil, Chile (Universidad de La Serena), and Spain (Centro de Estudios de Física del Cosmos de Aragón, CEFCA). We further acknowledge financial support from the São Paulo Research Foundation (FAPESP), Fundação de Amparo à Pesquisa do Estado do RS (FAPERGS), the Brazilian National Research Council (CNPq), the Coordination for the Improvement of Higher Education Personnel (CAPES), the Carlos Chagas Filho Rio de Janeiro State Research Foundation (FAPERJ), and the Brazilian Innovation Agency (FINEP). The authors who are members of the S-PLUS collaboration are grateful for the contributions from CTIO staff in helping in the construction, commissioning and maintenance of the T80-South telescope and camera. We are also indebted to Rene Laporte and INPE, as well as Keith Taylor, for their important contributions to the project. From CEFCA, we particularly would like to thank Antonio Marín-Franch for his invaluable contributions in the early phases of the project, David Cristóbal-Hornillos and his team for their help with the installation of the data reduction package jype version 0.9.9, César Íñiguez for providing 2D measurements of the filter transmissions, and all other staff members for their support with various aspects of the project. P.K.H. gratefully acknowledges the Fundação de Amparo à Pesquisa do Estado de São Paulo (FAPESP) for the support grant 2023/14272-4. LSJ acknowledges the support from CNPq (308994/2021-3)  and FAPESP (2011/51680-6). CMdO acknowledges the support from CNPq (307879/2025-9)  and FAPESP (2019/26492-3). LLN thanks Conselho Nacional de Desenvolvimento Científico e Tecnológico (CNPq) for granting the postdoctoral research fellowship 151798/2025-7. This research has made use of the SIMBAD database, operated at CDS, Strasbourg, France.
\end{acknowledgements}

%
%

\bibliographystyle{mnras} 
\bibliography{Haack}

\newpage

\appendix

\section{PCA}
\label{app:histoPCA}

\begin{figure}[h!]
    \centering
    \includegraphics[width=1\columnwidth]{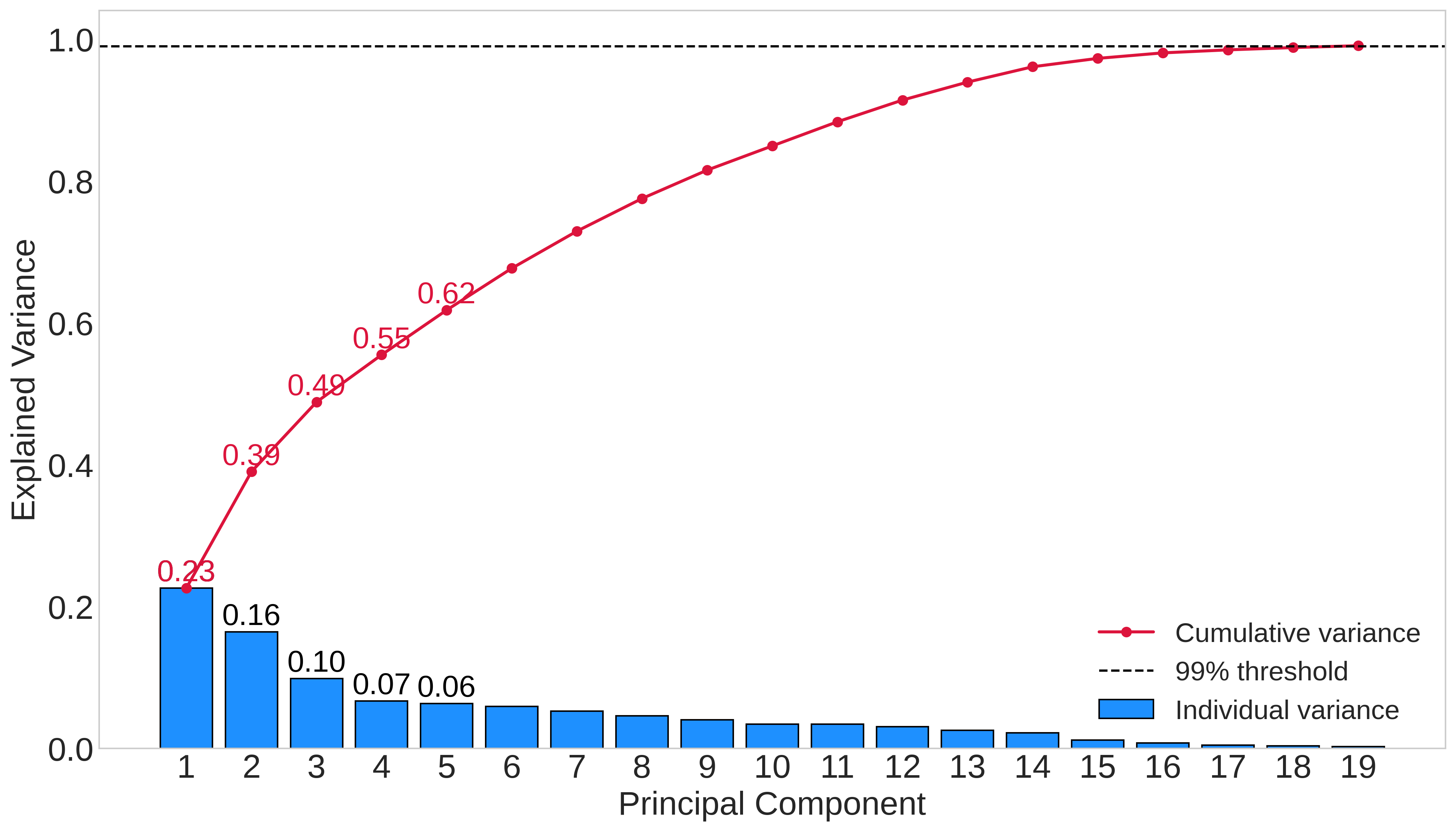}
    \caption{Histogram for the variance explained by each principal component.}
    \label{fig:histo_pca}
\end{figure}

\begin{figure}[h!]
    \centering
    \includegraphics[width=1\columnwidth]{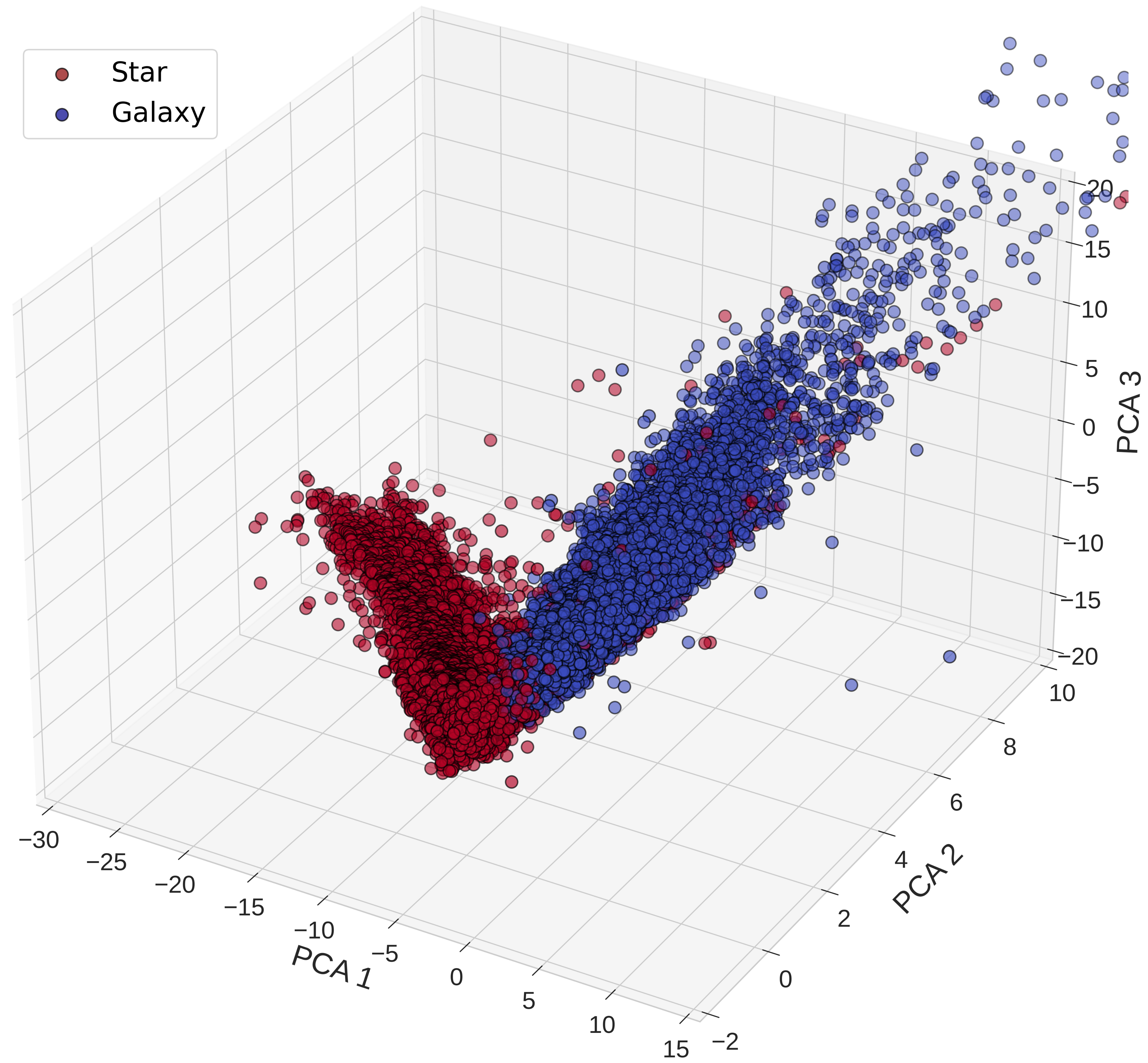}
    \caption{Star/galaxy separation in a 3D plot constructed with the three main components that contribute most to explaining the variance in the data.}
    \label{fig:PCA_3D}
\end{figure}

The histogram in the Figure\,\ref{fig:histo_pca} shows the variance explained by each principal component in the PCA analysis, highlighting the contribution of each component
to the dimensionality reduction. The red function represents the cumulative variance of the components until it explains 99$\%$ of the variance of the input data, a limit represented by the black-dashed horizontal line. Figure\,\ref{fig:PCA_3D} shows the star/galaxy separation in a 3D plot constructed with the three main components that contribute most to explaining the variance in the data.

\begin{figure*}[h!]
    \centering
    \includegraphics[width=1\linewidth]{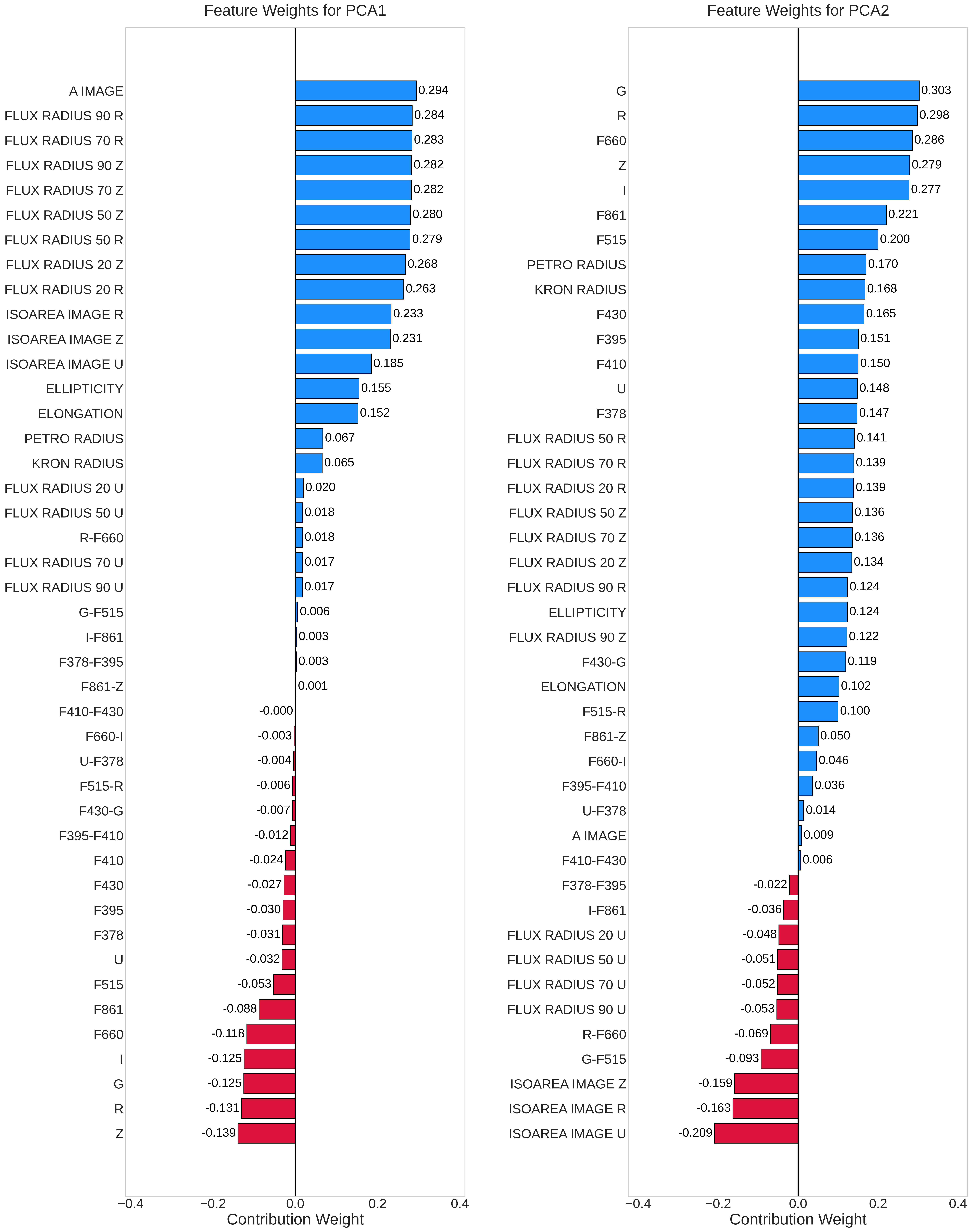}
    \caption{PCA1 and PCA2 feature weights.}
    \label{fig:pca_feauters}
\end{figure*}

The histograms in Figure\,\ref{fig:pca_feauters} visualize the contribution weights of each feature to the first two principal components (PCA1 and PCA2) of Figure\,\ref{fig:histo_pca}, displaying, from top to bottom, features sorted by absolute impact magnitude. Features with blue bars exhibit positive correlations that increase the component value, while red bars represent negative correlations that decrease it. The horizontal bar lengths quantify each feature's relative influence in defining the component's direction in the reduced-dimensional space, with labels indicating exact weight values positioned adjacent to bars for clear readability. This representation identifies which original variables most significantly shape the principal components' variance structure. For PCA1 the impact on star/galaxy separation is greater for features that have information on the size or geometry of the sources compared to features that provide photometric information. It should also be noted that the impact of these types of features is less evident for PCA2. 

\section{Galaxy sample with background problems}
\label{app:back_probl}

\begin{figure*}[h!]
    \centering
    \includegraphics[width=1\textwidth]{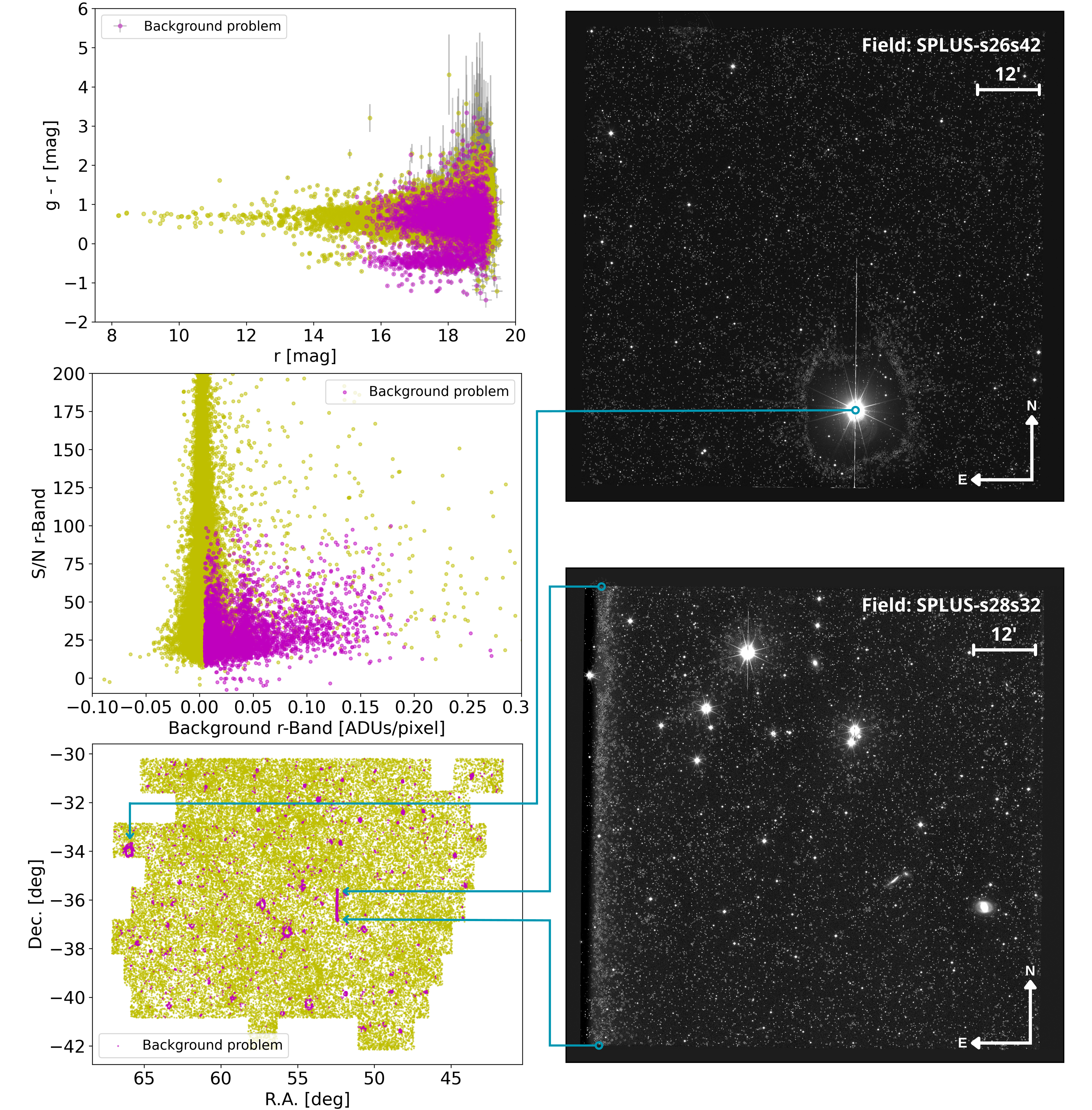}
    \caption{Identification and selection of problematic background galaxies.}
    \label{fig:background_problem}
\end{figure*}

In the catalog of 134,593 galaxies we have been able to identify atypical galaxy-patterned overdensities in some specific areas of the spatial distribution of the 106 S+FP fields. These galaxies survive due to two types of problems. One field (ID S-PLUS-s28s32; R.A.= +3:26:40,0 Dec= -36:11:27,5) presents problems in the broad-band images. On the other hand, the brightest saturated stars highly disturb the background measurement of SExtractor. Both are abnormal and extreme issues that affect the performance of the cleaning and classification algorithms. It is worth mentioning that the objects defining the atypical overdensities are galaxies and not spurious objects. As a consequence, we have decided to flag them in the catalog as {\it background problem}. To select them, we used the color-magnitud diagram shown in the top panel of Figure\,\ref{fig:background_problem}. There, a bluer sequence parallel to the red sequence defined by the early-type galaxies is evident. In addition, we identify that for values of the signal-to-noise ratio (S/N) in the r-band less than 100, there are galaxies displaying a wide range of values in the sky background measured by SExtractor (0.005 < BACKGROUND$\_$r < 0. 3). Those objects generate two vertical density columns and a cloud over the entire background range, as can be seen in the central panel of Figure\,\ref{fig:background_problem}. In addition, the vertical overdensity columns are populated with objects of apparently intermediate or large size (KRON$\_$RADIUS > 8). Galaxies selected with this criterion are highlighted in magenta in the spatial distribution shown in the lower panel of Figure\,\ref{fig:background_problem}, where it is evident that the "ring" shaped overdensities are linked to extreme cases of saturated stars disturbing the projected peripheral sky brightness. On the other hand, there is an elongated structure that corresponds to a problem in the image of the field ID S-PLUS-s28s32. Due to the aforementioned issues, the final catalog contains a total of 134,593 galaxies with 15,013 flaged objects due to background problems.

\clearpage
\section{Filter Set}
\label{app:filter}
\begin{figure*}[h!]
    \centering
    \includegraphics[width=1.0\textwidth]{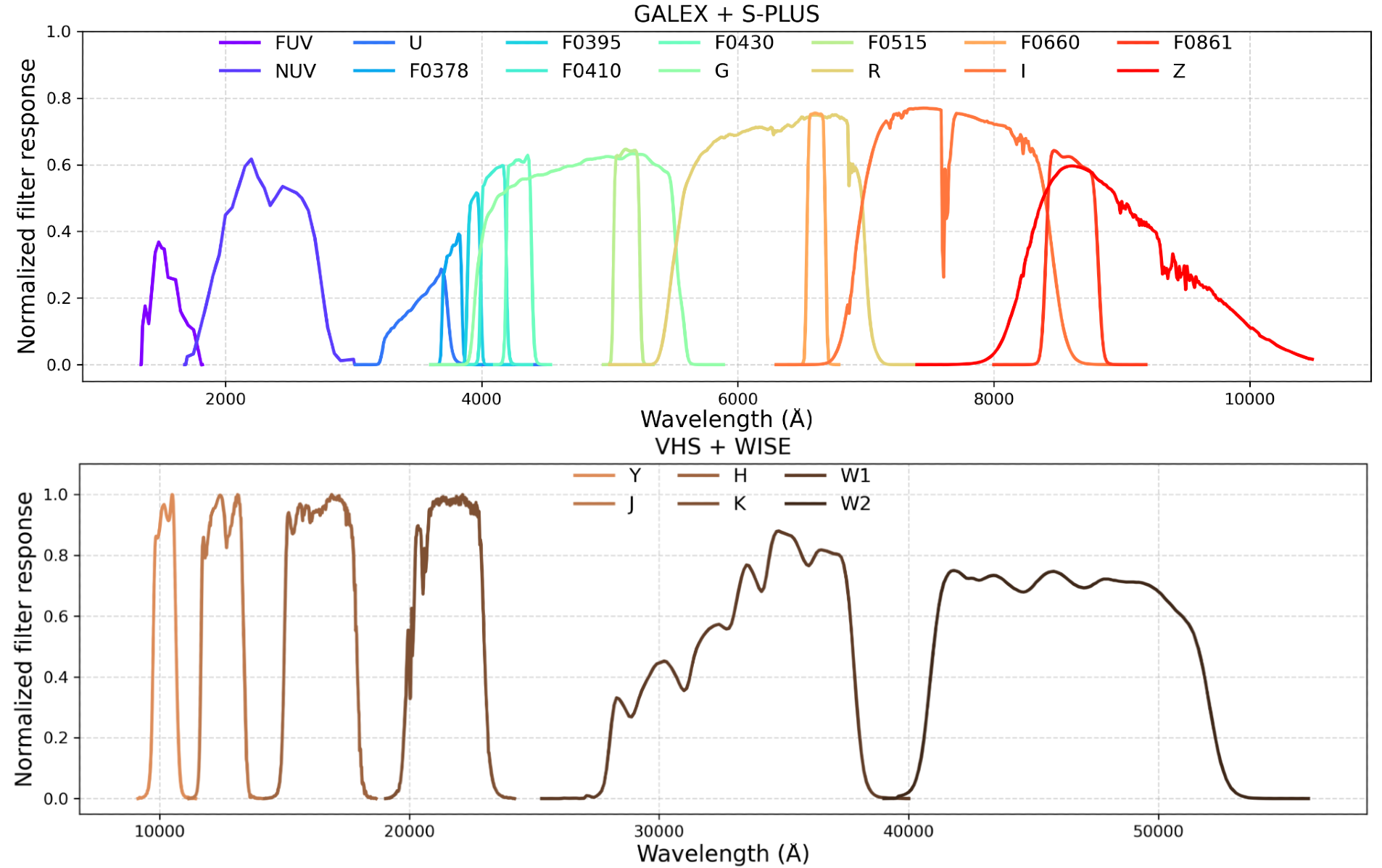}
    \caption{The combination of filters used and their respective normalized transmission curves. On the top panel, GALEX and S-PLUS filters. In the lower panel, VHS and AllWISE filters.}
    \label{fig:filter_set}
\end{figure*}

Further details on the photometric filter set used throughout the work are given here. The S-PLUS data have been combined with surveys in the UV (GALEX) and IR (VHS-VISTA and AllWISE). Table \ref{tab:filtros} lists the filters used, their central wavelength ($\lambda_c$), their respective width ($\Delta\lambda$) and the survey which each filter corresponds to. On the other hand, in Figure \ref{fig:filter_set}, the normalised transmission curves of all the filters that integrate the filter set are shown.

\begin{table}[ht]
\centering
\caption{Photometric filters sorted by central wavelength (UV to IR)}
\label{tab:filtros}
\begin{tabular}{llcc}
\toprule
\hline
\textbf{Filter} & $\boldsymbol{\lambda_c}$ (\textbf{\AA}) & $\boldsymbol{\Delta\lambda}$ (\textbf{\AA}) & \textbf{Survey} \\
\midrule
\hline
FUV      & 1539   & 443    & GALEX \\
NUV      & 2316   & 1066   & GALEX \\
u        & 3536   & 398    & S-PLUS \\
J0378    & 3770   & 168    & S-PLUS \\
J0395    & 3940   & 202    & S-PLUS \\
J0410    & 4094   & 208    & S-PLUS \\
J0430    & 4292   & 200    & S-PLUS \\
g        & 4751   & 1539   & S-PLUS \\
J0515    & 5133   & 202    & S-PLUS \\
r        & 6258   & 1479   & S-PLUS \\
J0660    & 6614   & 148    & S-PLUS \\
i        & 7690   & 1470   & S-PLUS \\
J0861    & 8611   & 408    & S-PLUS \\
z        & 8831   & 695    & S-PLUS \\
Y        & 10200  & 1200   & VHS-VISTA \\
J        & 12540  & 1620   & VHS-VISTA \\
H        & 16460  & 2900   & VHS-VISTA \\
Ks       & 21500  & 3200   & VHS-VISTA \\
W1       & 34000  & 6600   & AllWISE \\
W2       & 46000  & 10400  & AllWISE \\
W3       & 120000 & 55000  & AllWISE \\
W4       & 220000 & 41000  & AllWISE \\
\bottomrule
\end{tabular}
\end{table}

\newpage
\section{Odds}
\label{app:odds}
\begin{figure}[h!]
    \centering
    \includegraphics[width=1\columnwidth]{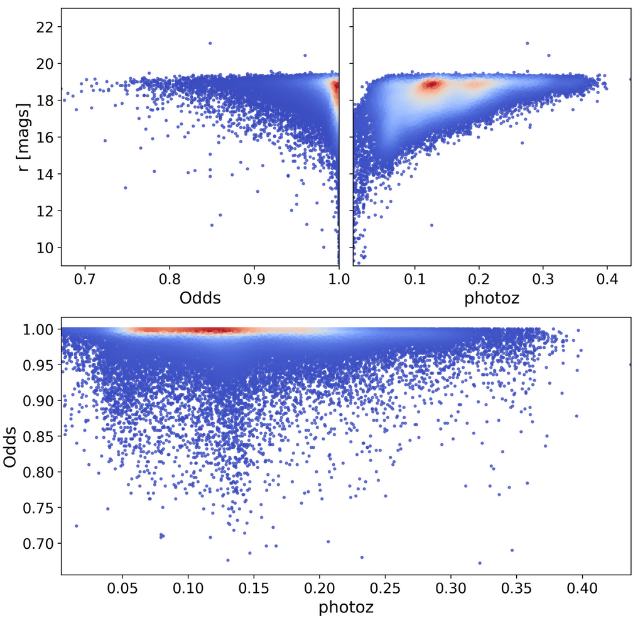}
    \caption{The relationships between $z_{phot}$, {\it Odds} and r AUTO magnitudes.}
    \label{fig:odds}
\end{figure}

The relationships found between $z_{phot}$, {\it Odds} and r-band magnitudes are presented here. As can be seen in Figure\,\ref{fig:odds}, in the {\it Odds} versus r-band magnitude plane (upper left panel), it is remarkable how all the computed {\it Odds} are above 0.7. It is worth remembering that the range of allowed values for {\it Odds} 0.7. It is 0.0 to 1.0, and higher {\it Odds} values imply that each source's PDF is narrower. It is noticeable that, for the brightest galaxies, the {\it Odds} are close to 1.0 while, at fainter magnitudes, especially reaching the limit of the photometric depth in the r-band ($\sim$ 19.5 mag), the {\it Odds} values are distributed from 0.7 to 1.0 concentrating mostly near 1.0.

In the upper right panel, the brightest galaxies (10 mag < r < 15 mag) are mostly found at the lowest $z_{phot}$, while galaxies with r > 15 mag, are distributed down to $z_{phot}$ < 0.4. This shows that only galaxies with r > 18 mag have 0.3 < $z_{phot}$ < 0.4. The distribution over the entire $z_{phot}$ range is not homogeneous, evidencing several overdensities, for example between 0.1 < $z_{phot}$ < 0.2.

In the lower panel where we show $z_{phot}$ versus {\it Odds}, the aforementioned overdensities are again observed at 0.1 < $z_{phot}$ < 0.2), and the {\it Odds} takes values between 0.7 and 1.0 over the entire range of $z_{phot}$. In this sense, there is no clear correlation between Odds and $z_{phot}$.

\section{Format of The Extragalactic Catalog}
\label{sec:format}

Tables\,\ref{tab:catalog_columns_1} and \ref{tab:catalog_columns_2} show the format of the extragalactic catalog, providing the name of each column, a description, the type of variable and the corresponding units, respectively. Both tables correspond to the same set of 119,580 galaxies (i.e., they share the same number of rows). Table\,\ref{tab:catalog_columns_1} gives the general properties, i.e. astrometric parameters (including right ascension and declination), geometrical parameters, additional information that has been used or is useful for the user, and property estimates that have been obtained in this work. Table\,\ref{tab:catalog_columns_2} gives photometric properties from both S-PLUS and external surveys (GALEX, VHS-VISTA, and AllWISE). Together they form the complete catalog, comprising 493 columns in total. To minimize the size of the displayed table and avoid repetition, '${filter}$' is a generic designation for any of the 12 S-PLUS bands.

\begin{table}[ht]
\centering
\caption{General properties}
\label{tab:catalog_columns_1}
\begin{tabular}{p{5cm}p{7cm}p{3cm}l}
\hline
\textbf{Column} & \textbf{Description} & \textbf{Type} & \textbf{Units} \\ 
\hline
ID & Unique identifier & integer & -- \\
Field & Observation S-PLUS field & string & -- \\
RUN & Run number (1, 2 or 3) & integer & -- \\
Warning & Quality warning flag & integer & -- \\
Name\_Literature & FLS name & string & -- \\
SIMBAD\_main\_id & SIMBAD identifier & string & -- \\
SIMBAD\_main\_type & SIMBAD object type & string & -- \\
SIMBAD\_redshift & Redshift from SIMBAD & float & -- \\
SIMBAD\_redshift\_err & Redshift error from SIMBAD & float & -- \\
data\_missing\_AUTO & Missing data flag & boolean & -- \\
n\_data\_missing & Number of missing data points & integer & -- \\
z\_phot & ML $z_{\text{phot}}$ & float & -- \\
Odds & Odds for $z_{\text{phot}}$ & float & -- \\
sigma\_68 & $\sigma_{68}$ for $z_{\text{phot}}$ & float & -- \\
log\_mass & ML stellar mass & float & $M/M_\odot$ \\
log\_SFR & ML star formation rate & float & $M_\odot/yr$ \\
D4000\_N & ML $D4000_{N}$ index & float & -- \\
Gal\_type & Quiescent, star forming or transition & string & -- \\
RA & Right ascension & float & degrees \\
DEC & Declination & float & degrees \\
X\_IMAGE & X position in image & float & pixels \\
Y\_IMAGE & Y position in image & float & pixels \\
THETA\_IMAGE & Orientation angle & float & degrees \\
ERRTHETA\_IMAGE & Orientation angle error & float & degrees \\
A\_IMAGE & Semi-major axis & float & pixels \\
ERRA\_IMAGE & Semi-major axis error & float & pixels \\
B\_IMAGE & Semi-minor axis & float & pixels \\
ERRB\_IMAGE & Semi-minor axis error & float & pixels \\
X\_WORLD & World X coordinate & float & degrees \\
Y\_WORLD & World Y coordinate & float & degrees \\
THETA\_WORLD & World orientation angle & float & degrees \\
ERRTHETA\_WORLD & World orientation angle error & float & degrees \\
A\_WORLD & World semi-major axis & float & degrees \\
ERRA\_WORLD & World semi-major axis error & float & degrees \\
B\_WORLD & World semi-minor axis & float & degrees \\
ERRB\_WORLD & World semi-minor axis error & float & degrees \\
ELONGATION & Elongation ratio (A/B) & float & -- \\
ELLIPTICITY & 1 - (B/A) & float & -- \\
KRON\_RADIUS & Kron radius in units of A or B  & float & -- \\
PETRO\_RADIUS & Petrosian radius in units of A or B  & float & -- \\
\hline
\end{tabular}
\end{table}

\clearpage

\begin{table}[ht]
\centering
\caption{Photometric properties}
\label{tab:catalog_columns_2}
\begin{tabular}{p{5cm}p{7cm}p{3cm}l}
\hline
\textbf{Column} & \textbf{Description} & \textbf{Type} & \textbf{Units} \\ 
\hline
FLUX\_AUTO\_\{filter\} & Auto flux & float & erg/s/cm $^2$/\r{A} \\
FLUXERR\_AUTO\_\{filter\} & Auto flux error & float & erg/s/cm $^2$/\r{A} \\
\{filter\}\_AUTO & Auto magnitude & float & mag \\
e\_\{filter\}\_AUTO & Auto magnitude error & float & mag \\
FLUX\_ISO\_\{filter\} & Isophotal flux & float & erg/s/cm $^2$/\r{A} \\
FLUXERR\_ISO\_\{filter\} & Isophotal flux error & float & erg/s/cm $^2$/\r{A} \\
\{filter\}\_ISO & Isophotal magnitude & float & mag \\
e\_\{filter\}\_ISO & Isophotal magnitude error & float & mag \\
FLUX\_PETRO\_\{filter\} & Petrosian flux & float & erg/s/cm $^2$/\r{A} \\
FLUXERR\_PETRO\_\{filter\} & Petrosian flux error & float & erg/s/cm $^2$/\r{A} \\
\{filter\}\_PETRO & Petrosian magnitude & float & mag \\
e\_\{filter\}\_PETRO & Petrosian magnitude error & float & mag \\
\{filter\}\_APER\_3 & 3arcsec aperture magnitude & float & mag \\
\{filter\}\_APER\_6 & 6arcsec aperture magnitude & float & mag \\
e\_\{filter\}\_APER\_3 & 3arcsec aperture magnitude error & float & mag \\
e\_\{filter\}\_APER\_6 & 6arcsec aperture magnitude error & float & mag \\
FLUX\_APER\_3\_\{filter\} & 3arcsec aperture flux & float & erg/s/cm $^2$/\r{A} \\
FLUX\_APER\_6\_\{filter\} & 6arcsec aperture flux & float & erg/s/cm $^2$/\r{A} \\
FLUXERR\_APER\_3\_\{filter\} & 3arcsec aperture flux error & float & erg/s/cm $^2$/\r{A} \\
FLUXERR\_APER\_6\_\{filter\} & 6arcsec aperture flux error & float & erg/s/cm $^2$/\r{A} \\
FLAGS\_\{filter\} & Quality flags & integer & -- \\
FWHM\_IMAGE\_\{filter\} & FWHM in image coordinates & float & pixels \\
FWHM\_WORLD\_\{filter\} & FWHM in world coordinates & float & arcsec \\
ISOAREA\_IMAGE\_\{filter\} & Isophotal area in image  & float & pixels$^2$ \\
ISOAREA\_WORLD\_\{filter\} & Isophotal area in sky & float & arcsec$^2$ \\
FLUX\_RADIUS\_20\_\{filter\} & Radius enclosing 20\%  of the total flux & float & pixels \\
FLUX\_RADIUS\_50\_\{filter\} & Radius enclosing 50\%  of the total flux & float & pixels \\
FLUX\_RADIUS\_70\_\{filter\} & Radius enclosing 70\%  of the total flux & float & pixels \\
FLUX\_RADIUS\_90\_\{filter\} & Radius enclosing 90\%  of the total flux & float & pixels \\
FLUX\_MAX\_\{filter\} & Maximum flux & float & erg/s/cm $^2$/\r{A} \\
SNR\_WIN\_\{filter\} & Windowed signal-to-noise ratio & float & -- \\
MU\_THRESHOLD\_\{filter\} & Surface brightness threshold &  instrumental \\
THRESHOLD\_\{filter\} & Detection threshold & float &  instrumental \\
MU\_MAX\_\{filter\} & Maximum surface brightness & float &  instrumental \\
CLASS\_STAR\_\{filter\} & Stellarity index & float & -- \\
BACKGROUND\_\{filter\} & Background level & float &  instrumental \\
FUVmag & FUVmag Auto magnitude & float & mag \\
e\_FUVmag & FUVmag Auto magnitude error & float & mag \\
NUVmag & NUVmag Auto magnitude & float & mag \\
e\_NUVmag & NUVmag Auto magnitude error & float & mag \\
Ypmag & Ypmag Petrosian magnitude & float & mag \\
e\_Ypmag & Ypmag Petrosian magnitude error & float & mag \\
Jpmag & Jpmag Petrosian magnitude & float & mag \\
e\_Jpmag & Jpmag Petrosian magnitude error & float & mag \\
Hpmag & Hpmag Petrosian magnitude & float & mag \\
e\_Hpmag & Hpmag Petrosian magnitude error & float & mag \\
Kspmag & Kspmag Petrosian magnitude & float & mag \\
e\_Kspmag & Kspmag Petrosian magnitude error & float & mag \\
W1mag & W1mag magnitude in 8.25" aperture & float & mag \\
e\_W1mag & W1mag magnitude error & float & mag \\
W2mag & W2mag magnitude in 8.25" aperture & float & mag \\
e\_W2mag & W2mag magnitude error & float & mag \\
W3mag & W3mag magnitude in 8.25" aperture & float & mag \\
e\_W3mag & W3mag magnitude error & float & mag \\
W4mag & W4mag magnitude in 8.25" aperture & float & mag \\
e\_W4mag & W4mag magnitude error & float & mag \\

\hline
\end{tabular}
\end{table}

\end{document}